\documentclass[modern]{aastex63}
\usepackage{epsfig}
\usepackage{amsmath}
\usepackage{amstext}
\usepackage{amssymb}
\usepackage{color}

\def \micron{$\mu$m}

\def \iras {{\it IRAS}}
\def \spitzer {{\it Spitzer}}
\def \iue {{\it IUE}}

\def \ebv {$E(B-V)$}
\def \bump {2175~\AA}
\def \abump {$A(2175)$}

\accepted{November 8, 2021}
\submitjournal{ApJ}

\shorttitle{mid-IR emission and UV extinction}

\shortauthors{Massa, Gordon \& Fitzpatrick}

\begin{document}

\title{Relations between mid-IR dust emission and UV extinction}

\correspondingauthor{Derck Massa}
\email{dmassa@spacescience.org}

\author[0000-0002-9139-2964]{Derck Massa}
\affil{Space Science Institute, 4750 Walnut Street, Suite 205, Boulder, CO
80301, USA}

\author[0000-0001-5340-6774]{Karl D.\ Gordon}
\affiliation{Space Telescope Science Institute, 3700 San Martin
  Drive, Baltimore, MD, 21218, USA}
\affiliation{Sterrenkundig Observatorium, Universiteit Gent,
  Gent, Belgium}

\author[0000-0002-2371-5477]{E.\ L.\ Fitzpatrick}
\affiliation{Department of Astronomy \& Astrophysics, Villanova University,
800 Lancaster Avenue, Villanova, PA 19085, USA}

\begin{abstract}

We analyze low resolution \spitzer\ infrared (IR) 5-14~\micron\ spectra
of the diffuse emission toward a carefully selected sample of stars.  The
sample is composed of sight lines toward stars that have well determined
ultraviolet (UV) extinction curves and which are shown to lie beyond
effectively all of the extinguishing and emitting dust along their lines
of sight.  Our sample includes sight lines whose UV curve extinction
curves exhibit a wide range of curve morphology and which sample a
variety of interstellar environments.  As a result, this unique sample
enabled us to study the connection between the extinction and emission
properties of the same grains, and to examine their response to
different physical environments.  We quantify the emission features in
terms of the PAH model given by \citet{dl07} and a set on additional
features, not known to be related to PAH emission.  We compare the
intensities of the different features in the \spitzer\ mid-IR spectra
with the \citet{fm07} parameters which describe the shapes of UV to
near-IR extinction curves.  Our primary result is that there is a
strong correlation between the area of the \bump\ UV bump in the
extinction curves of the program stars and the strengths of the major
PAH emission features in the mid-IR spectra for the same lines of sight.
\end{abstract}
\vspace{0.5in}


\section{Introduction}
Interstellar dust is a ubiquitous component of the interstellar medium
(ISM). It plays a major role in determining the energy balance of the
ISM, acts as a sink for metals, a site for molecule formation, and may
even account for the confusing deuterium abundance results
\citep{2006ApJ...647.1106L}. Consequently, determining its composition is
of fundamental importance to astrophysics.  Two of the primary tools for
studying ISM dust composition are its extinction (absorption and scattering)
and infrared emission.

Diffuse ISM extinction curves exhibit only a few strong diagnostic
features: 1) their overall shapes, which are largely determined by the
grain size and composition distributions, 2) the \bump\ bump, 
3) the 3.4~\micron\ feature, and 4) a strong feature centered near
10~\micron\ \citep{2004ApJ...609..826K, 2011ApJ...740...93S}.  There is also
a broad, weak feature near 18~\micron\ \citep{2011A&A...526A.152V}, which is
thought to be due to O$-$Si$-$O bending, and three or more broad, weak
features in the optical \citep{2020ApJ...891...67M}, two of which are
related to the \bump\ UV bump, and numerous, narrow diffuse interstellar
bands in the optical \citep[e.g.][]{2019ApJ...878..151F}, whose carriers
remain largely unidentified.

In contrast, ISM dust has a rich emission spectrum in the mid-IR.  In the
wavelength range $3 \le \lambda \le 20$~\micron, the emission is dominated by
strong emission bands \citep{2004ApJS..154..309W}, which are ascribed to
material that includes different C$-$C and C$-$H stretching and bending modes.
All of these features appear on top of a broad continuum, whose shape appears
to depend upon line of sight conditions.  The features between $3 \le \lambda
\le 15$~\micron\ have received the most attention. The most promising
identification of the material responsible for these features is one composed
of large Polycyclic Aromatic Hydrocarbons, PAHs, \citep[e.g.,][]{1985ApJ...290L..25A,
dl07}.  Not only do the bands coincide with laboratory and theoretical PAH
spectra \citep[e.g.,][]{2001ApJ...556..501B}, but the response of the
individual band strengths and their ratios in classes of objects with
specific environments reinforces these identifications \citep[see, e.g.,][for a
review]{2004ApJ...617L..65P}.

It has long been suspected that the that the \bump\ absorption band is
related to PAHs.  The connection between PAHs and UV extinction has
been studied for a number of years, in the laboratory, with theoretical
calculations, and observationally with direct measurement of UV extinction
curves.  Laboratory measurements and theoretical calculations of specific
PAH molecules show sharp features in the ultraviolet
\citep{1992ApJ...395..301S, 1995P&SS...43.1165S, 2004A&A...426..105M,
2020ApJS..251...22M}, yet all known UV extinction curves are extremely
smooth, and only show the broad \bump\ bump \citep{Fitzpatrick86,
Fitzpatrick88, Fitzpatrick90, 2003ApJ...592..947C, Valencic04}.
Fortunately, laboratory measurements and theoretical calculations of PAH
populations with a wide size distribution display smooth extinction
including a broad feature near \bump\ \citep{1992ApJ...393L..79J,
2004A&A...426..105M, 2007A&A...462..627M, 2008A&A...489.1183M,
2010ApJ...712L..16S, 2011ApJ...742....2S, 2012A&A...540A.110S}. While the
laboratory spectra of PAH populations do not exactly match the observed UV
extinction curves, the correspondence suggests PAHs may contribute to a
portion of the UV extinction.

The most direct way to relate the \bump\ feature to PAHs, would be to
compare its extinction strength to that of mid-IR PAH features for the
same line of sight.  However, due to the enormous opacity of dust in the
UV, the line of sight with the largest $A(V)$ for which a \bump\
extinction bump has been measured is toward HD~283809, which has $A(V) ~
6$~mag \citep{2003ApJ...592..947C}, and it is doubtful that significantly
larger values will be reached with current telescopes.  In contrast, mid-IR
PAH features have only been definitively measured in sight lines with $A(V)
\sim 30$~mag \citep{2013ApJ...770...78C}.  More recently, a possible
detection of PAH features toward Cyg OB-12 (with an $A(V) \sim 10$~mag)
has been reported \citep{2020ApJ...895...38H}, but this has been
questioned by \citet{2021NatAs...5...78P}.  Thus, direct comparison of
\bump\ and mid-IR {\em extinction} is not feasible with current
observational capabilities.

While absorption by PAH features is quite weak, emission is much stronger
and provides a way forward.  One solution is to measure UV extinction and
IR emission from the same sightlines.  \citet{1993A&A...275..549J} combined
\iue\ extinction curves with \iras\ observations for comparing the UV and
IR properties along the same lines of sight and found that the UV bump
strength and 12~\micron\ emission were correlated.  Further,
\citet{2002A&A...382.1042V} used {\it Orbiting Infrared Observatory, ISO,}\
data to show a general trend of ``bump strength'' with the ratio of the
6.2 to 11.2~\micron\ line strengths, in the sense that as global bump
strength weakens in the series: Milky, LMC (no 30 Dor), LMC (30 Dor) and
SMC, the IR band ratios do too.  While this is not a quantitative relation
and relies on comparing regions with very different compositions and
environments, it does suggest that bump strength and PAH emission are
related at some level.  More recently, \citep{2017ApJ...836..173B} combined
\iue\ extinction measurements with \spitzer/IRS spectroscopy and find
correlate shifts between the \bump\ bump central wavelength and similar
shifts in two of the mid-infrared PAH features.

Our study builds on these results by using a carefully selected sample of
stars beyond the Galactic dust layer.  In this way, we can be certain that
{\em the extinction and line of sight emission originate from the same grain
population}.  This allows us to test, among other things, the conjecture
that all or part of the \bump\ extinction is due to PAHs, as claimed in
some ISM dust models \citep[e.g.,][]{2001ApJ...554..778L, 2004ApJS..152..211Z,
2007ApJ...663..866D}.  In these models, absorption of UV radiation by the
\bump\ bump is a major source for heating the PAHs, which then emit their
energy through their IR bands.  As a result, we might expect a relationship
between the strength of the \bump\ bump and the intensity of the PAH bands.

Section~\ref{sec:sample} describes how we selected our sample of sight lines
and shows the extinction curves for the sample of stars.
Section~\ref{sec:data} provides an overview of the \spitzer\ IRS
observations, how the spectra were obtained and their reduction.
Section~\ref{sec:errors}, discusses the derivation of the errors and
systematic effects associated with the spectra.  Section~\ref{sec:analysis}
presents our analysis of the spectra, and the model used to extract
quantitative measurements of the major spectral features.
Section~\ref{sec:results} gives our results and examines them for
correlations with each other and ancillary quantities.  Finally,
section~\ref{sec:discussion} provides a summary and brief discussion of our
findings.

\section{The Sample \label{sec:sample}}
Our sample was selected from the 332 stars in the \citet{fm07} UV extinction
atlas and the more lightly reddened stars in \citet{fm05}.  For each of
these stars, we compared the extinction color excesses listed in the catalog,
$E(B-V)$, to the line of sight ``emission color excesses'', $E(B-V)_{IR}$.
The latter are the color excesses calculated by \citet{1998ApJ...500..525S}
from temperature corrected \iras\ 100~\micron\ intensities  of dust along
the line of sight
\footnote{\citet{1998ApJ...500..525S} calibrated their relation between
emission and extinction using the colors of elliptical galaxies to measure
the reddening per unit flux density of 100 \micron\ emission.
}.

\begin{figure}
\includegraphics[width=0.9\linewidth]{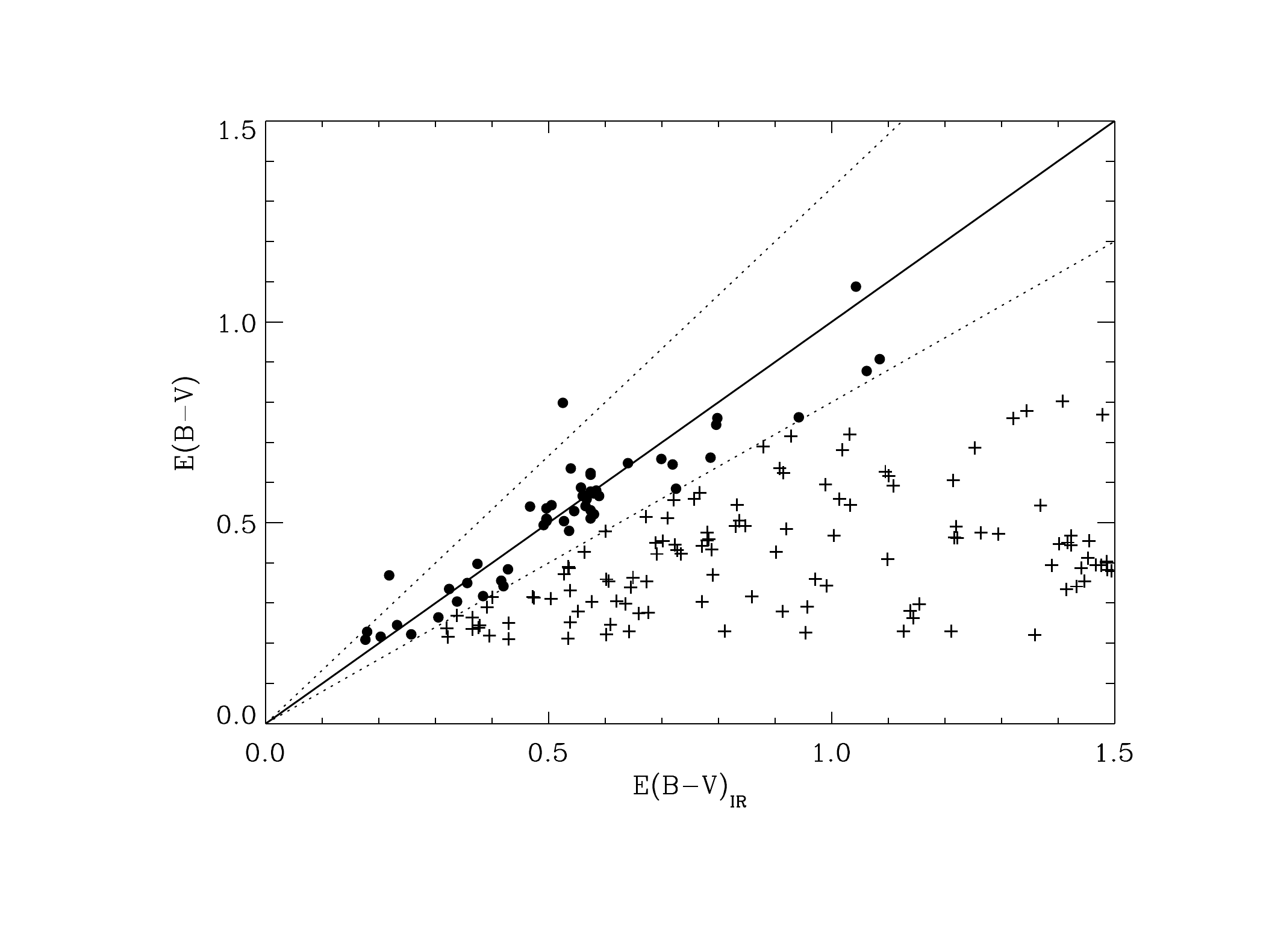}\vspace{-0.5in}
\vspace{-.2in}\caption{Color excesses derived from the MK type,
$E(B-V)_{\rm MK}$, versus the color excesses determined from {\it IRAS}\/
colors by \citet{1998ApJ...500..525S}, $E(B-V)_{\rm IRAS}$. The solid line
denotes exact agreement, and the two dotted lines are for $E(B-V) = (1 \pm
0.1)E(B-V)_{\rm IRAS}$.  Stars within the region bounded by the dotted lines
are plotted as solid points.  These stars constitute our initial (unculled)
sample.}
\label{fig:ebv-ebv}
\end{figure}

\begin{figure}
 \begin{center}
  \includegraphics[width=\linewidth]{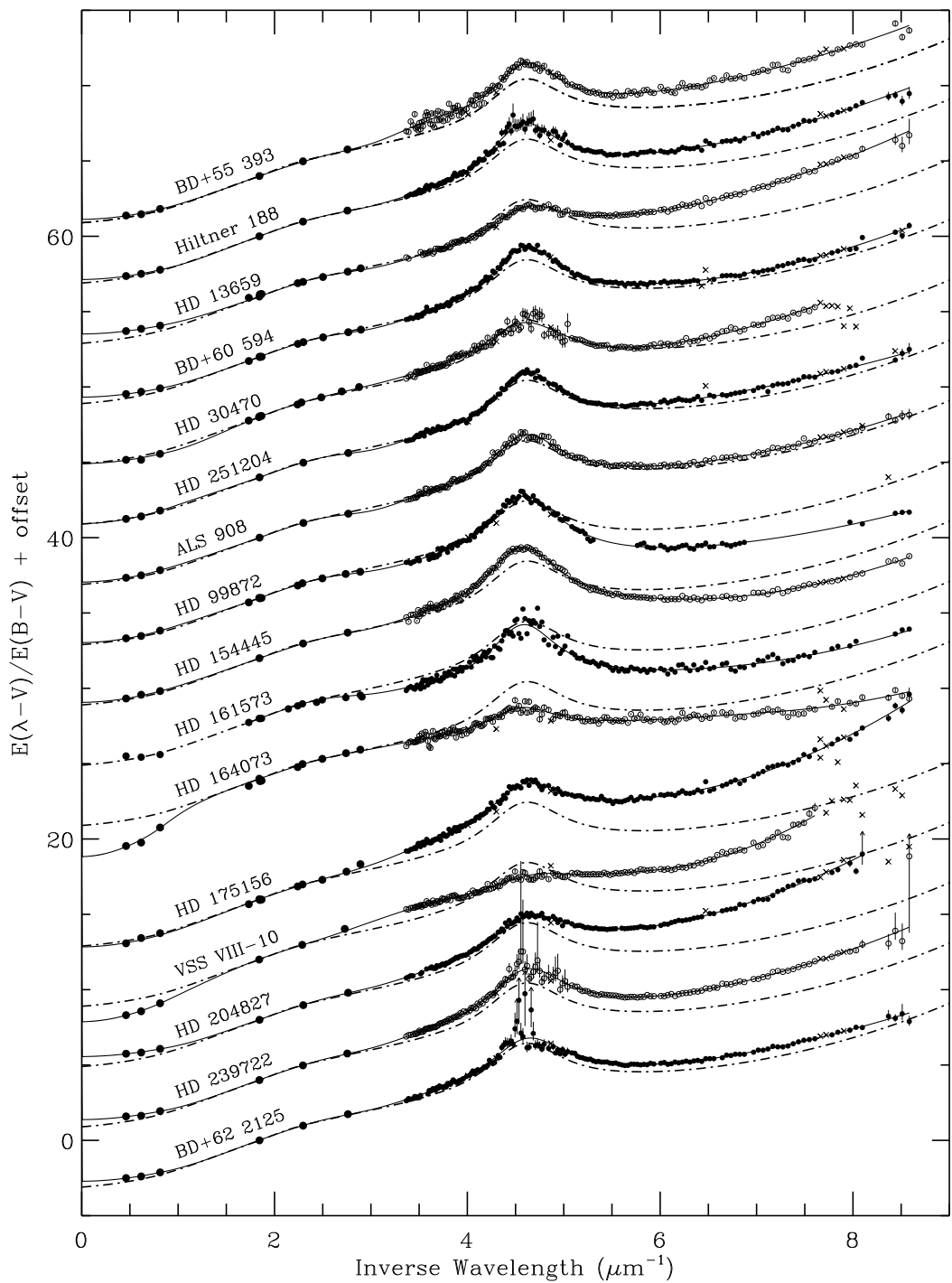}
  \end{center}
  \vspace{-0.2in}
  \caption{Extinction curves for the sample \citep[from][]{fm07}.  Each
  curve offset for display.  Also shown are analytic fits to   the data
  (solid curves), following the FM07 formulation, and their sample mean
  curve, for 332 Galactic stars, to demonstrate the range of UV
  properties.}\label{fig:uvcurves}
\end{figure}

The two color excesses are shown in Figure~\ref{fig:ebv-ebv}.  Only stars
for which these two measures of dust column density agreed to within $\pm
10$\% were included for further consideration.  It is clear that for these
stars, the bulk of the emitting and extincting dust lies in front of them.
The sample was further restricted to sight lines with a range of UV
extinction curve properties and environments.  Table~\ref{tab:sample}
lists our final sample of 16 stars along with the properties of their lines
of sight, such as cluster membership and association with dark clouds.
All of the distances listed are from {\em Gaia}\ \citep{2018yCat.1345....0G},
except for HD~99872 and HD~175156, which are from {\em Hipparcos}
\citep{2007A&A...474..653V}.  No parallax is available for VSS VIII-10.
Figure~\ref{fig:uvcurves} shows the range of UV curve types sampled by the
program stars.

\begin{table}[h]
\caption{Program Stars}
\begin{center}
\begin{tabular}{llcrrrrrrl}
\hline
Star        & Sp Ty & $E(B-V)$ & $R(V)$ & $l$ & $b$  & \multicolumn{2}{c}{Ecliptic}
            & Distance & Environment   \\
            &       & mag      &    & deg & deg  & long  & lat &  \multicolumn{1}{c}{kpc} \\
            \hline
BD+55$^\circ$393   & B1 V      & 0.26 & 2.84 & 130.3 &  -6.0 &   48.0 &  41.6 & $2.49 \pm 0.26$ & below h\&$\chi$\ Per \\
Hiltner 188 & B1 V      & 0.66 & 2.86 & 131.2 &  -1.8 &   52.7 &  44.1 & $4.04 \pm 0.53$ & above h\&$\chi$\ Per \\
HD 13659    & B1 Ib     & 0.80 & 2.48 & 134.2 &  -4.1 &   53.5 &  40.4 & $2.24 \pm 0.20$ & near h\&$\chi$\ Per \\
BD+60$^\circ$594   & A0 V      & 0.65 & 2.66 & 137.4 &   2.1 &   62.4 &  42.3 & $2.16 \pm 0.15$ & IC 1848  \\
HD 30470    & B2 V      & 0.34 & 3.26 & 187.7 & -21.1 &   72.1 & -11.4 & $0.79 \pm 0.06$ & NGC 1662 \\
HD 251204   & B0 V      & 0.76 & 3.07 & 187.0 &   1.0 &   91.2 &  -0.0 & $6.73 \pm 2.41$ & --       \\
ALS 908     & O9 V      & 0.65 & 2.99 & 245.8 &   0.6 &  130.7 & -47.9 & $8.98 \pm 2.24$ & Rup 44   \\
HD 99872    & B3 V      & 0.32 & 2.99 & 296.7 & -10.6 &  228.8 & -63.0 & $0.24 \pm 0.02$ & double   \\
HD 154445   & B1 V      & 0.40 & 2.95 &  19.3 &  22.9 &  255.3 &  21.9 & $0.27 \pm 0.01$ & --       \\
HD 161573   & B3 IV     & 0.20 & 3.10 &  30.4 &  17.0 &  266.1 &  28.9 & $0.27 \pm 0.01$ & IC 4665  \\
HD 164073   & B3 III-IV & 0.21 & 5.34 & 344.2 & -12.6 &  270.4 & -25.4 & $0.61 \pm 0.02$ & --       \\
HD 175156   & B5 III    & 0.35 & 3.13 &  19.3 &  -7.8 &  283.3 &   7.2 & $0.43 \pm 0.04$ & --       \\
VSS VIII-10 & B8 V      & 0.74 & 4.18 & 359.5 & -18.3 &  282.9 & -14.8 & \multicolumn{1}{c}{--} & Cor Aust \\
HD 204827   & B0 V      & 1.09 & 2.46 &  99.2 &   5.6 &    6.8 &  65.6 & $1.29 \pm 0.16$ & Tr 37    \\  
HD 239722   & B2 IV     & 0.88 & 2.61 & 100.3 &   5.1 &    9.3 &  65.0 & $0.78 \pm 0.02$ & Tr 37    \\
BD+62$^\circ$2125  & B1 V      & 0.91 & 2.71 & 110.1 &   3.5 &   28.9 &  60.7 & $0.77 \pm 0.02$ & Cep OB3  \\ 
\hline
\end{tabular}\label{tab:sample}
\end{center}
\end{table}

\clearpage

\clearpage

\section{The Data\label{sec:data}}

\subsection{The observations}
We obtained \spitzer\ IRS SL1 and SL2 spectra of the dust near the 16
program stars.  The regions parallel and perpendicular to the slit were mapped.
This furnished two distinct advantages.  First, it allowed us to search
for small scale spatial variability in the background.  Second, it occulted
the bright source, and eliminated the influence of scattered light on the
background.  We selected the mapping pointings to be perpendicular to the
slit so that they covered a range roughly comparable to the slit width.
This allowed us to test for variability over a roughly square area.

We obtained three independent samples of the background along the slit for
the mapping positions off of the source.  Overall, our short wavelength
observations gave us 9 independent spatial positions (3 slit positions, and
3 positions along the slit), with the source in the center position and 8
positions surrounding it.  The spatial regions covered by the slits on the
sky are illustrated in Fig.~\ref{fig:wavecut}.  In the figure, each
observation is represented by a set of diagonal lines, and regions where the
same portion of the sky is observed by different slits appears cross-hatched.

\subsection{The spectra}
We extracted short wavelength, low resolution (SL) spectra, which are
divided into two segments: the first order (SL1), which covers 7.46 to 14.29
\micron, and second order (SL2), which spans 5.13 to 7.60 \micron.

\begin{figure}
\begin{center}
\includegraphics[width=0.6\linewidth]{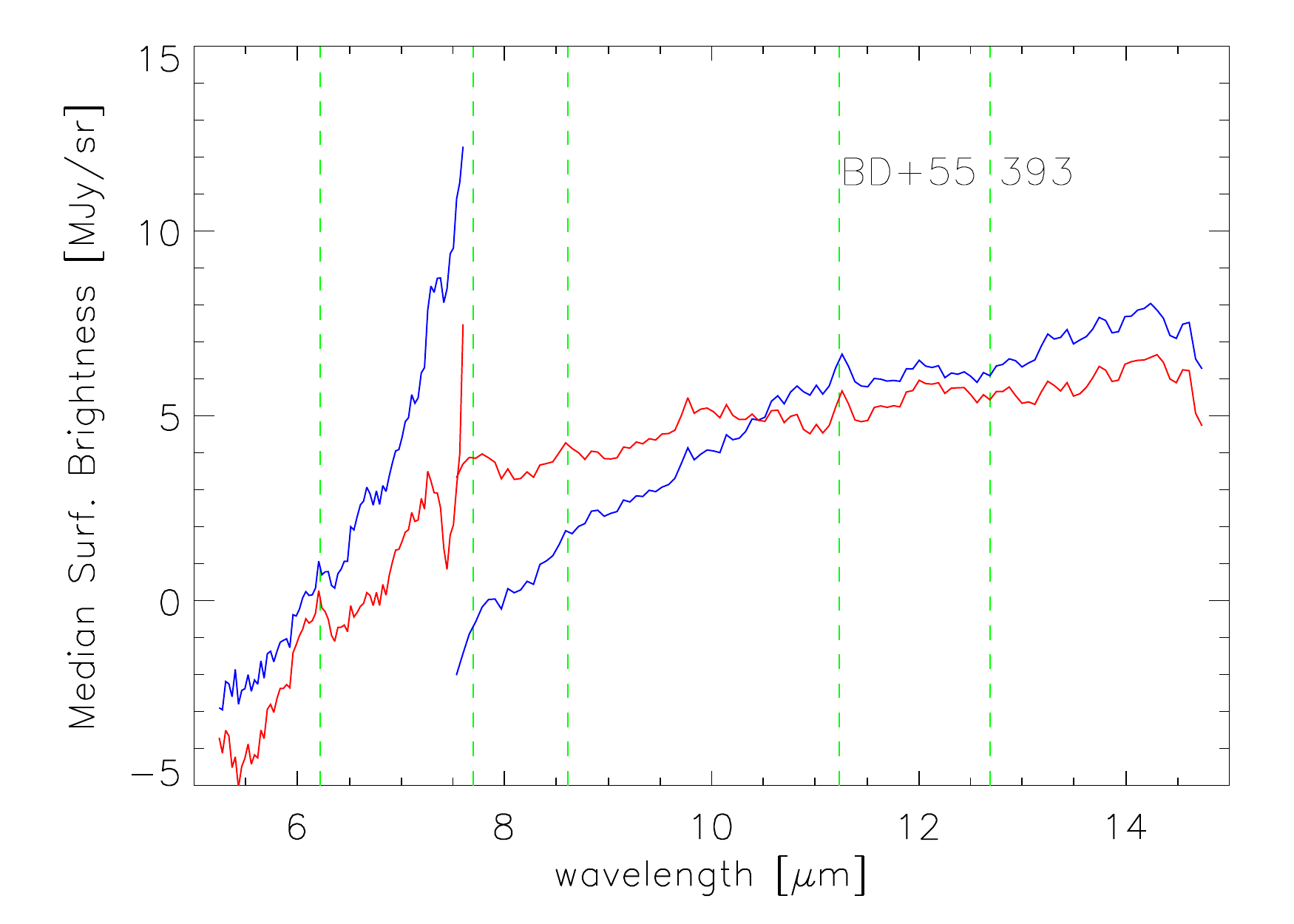}
\end{center}
\vspace{-0.2in}
\caption{The effect of the "dark settle" correction of the SL data is shown
for the region around BD+55$^\circ$~393.  The blue lines give the extracted
SL spectra before correction and the red after correction.  The green dashed
vertical lines show the locations of the of the strong aromatic features.
\label{fig:origvcorr}}
\end{figure}

The processing of the \spitzer\ SL IRS data started with the basic
calibrated data (BCD) images that were downloaded from the \spitzer\ archive
\citep[see][]{2004ApJS..154...18H}.  We applied a correction for the "dark
settle" issue using the inter-order regions \citep{2012ApJ...744...20S}.
This significantly improved the agreement in the overlap region between SL1
and SL2 orders as is illustrated for one of our targets in Fig.~\ref{fig:origvcorr}.

\begin{figure}
\begin{center}
  \includegraphics[width=0.6\linewidth]{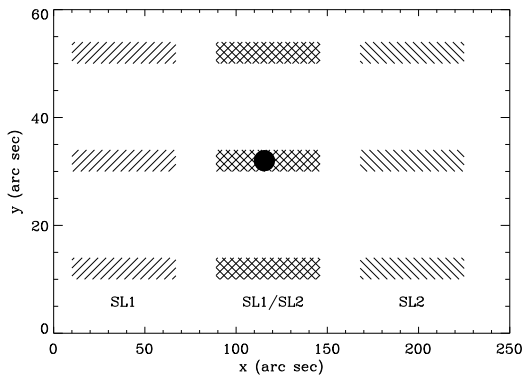}
\end{center}
\vspace{-2.2in}
\caption{A wavelength slice of a SL1 spectral cube showing the spatial
coverage of the 6 pointings, which cover 9 locations on the sky.  Eight
of the regions do not include the target star and were used to measure the
median surface brightness and its variations.
\label{fig:wavecut}}
\end{figure}

Spectral cubes were created using the CUBISM package \citep{2007PASP..119.1133S}.
CUBISM constructs spectral cubes for measurements of extended sources using
\spitzer\ IRS slit spectroscopy.  Spectral cubes were constructed from the
observations for each target for each spectral order.  This resulted in two
spectral cubes for each target, one for each of the orders.
The disjoint
spatial regions covered by the spectral cubes are illustrated in
Fig.~\ref{fig:wavecut}.

The median surface brightness at each wavelength was measured for each pixel
on the sky from the spectral cubes after masking the location the target star.
This results in approximately 175 independent spectra, which were used to
compute the error covariance matrices needed to quantify the correlated
errors between different wavelengths in the final, extracted spectra.

\section{Errors and Systematic Effects\label{sec:errors}}
\begin{figure}
\includegraphics[width=\linewidth]{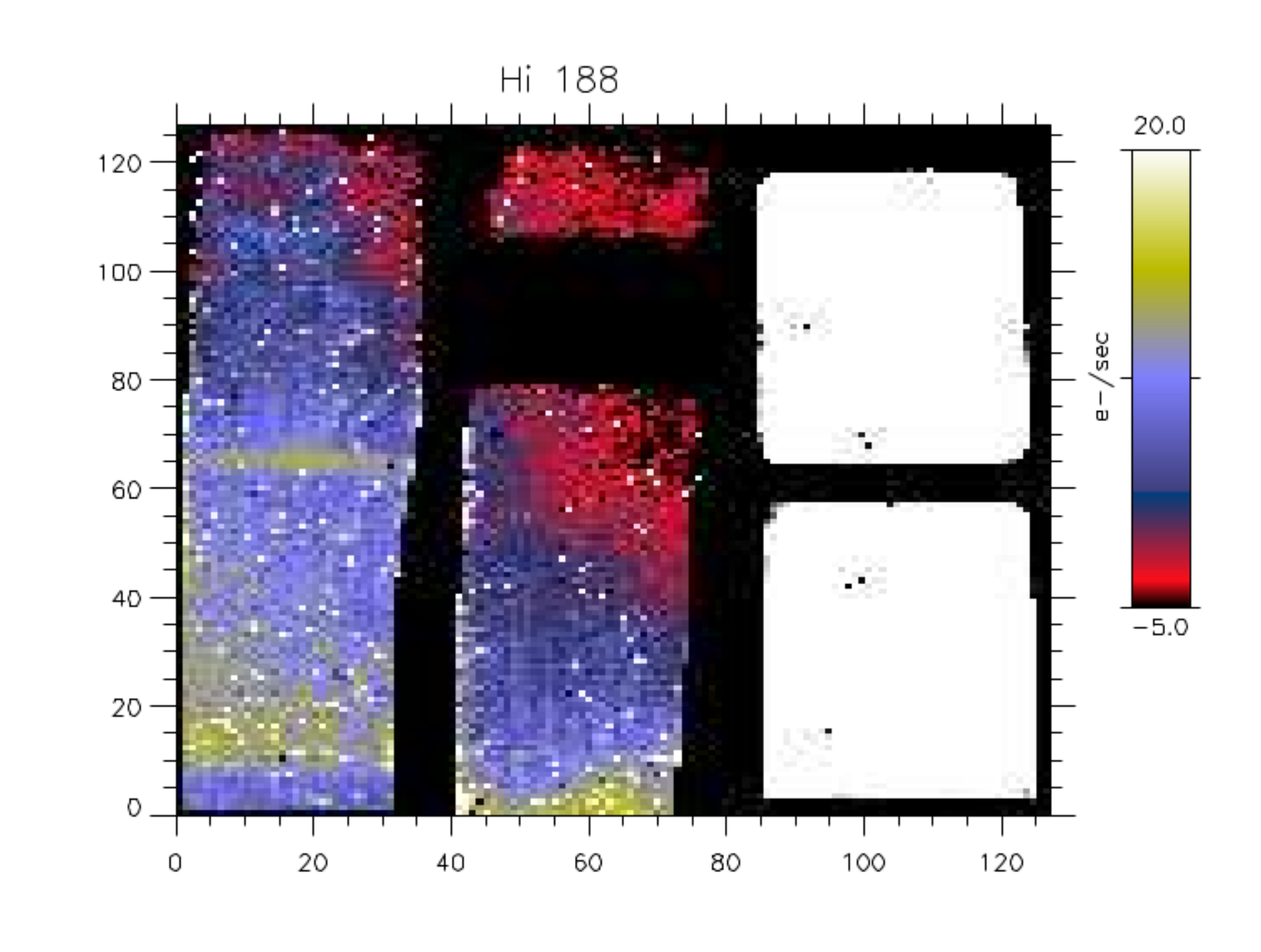}

\vspace{-0.3in}
\caption{Image of the detector for a diffuse background
observation near Hiltner~188.  The 11.3 \micron\ feature is clearly visible
in the SL1 spectrum on the left side of the image near $y=65$, $0 \leq x
\leq 30$.  The highly variable SL2 background is also apparent as the change
in intensity across the SL2 spectrum which occupies $40 \lesssim x \lesssim
70$}.\label{fig:image}
\end{figure}

The errors affecting the data arise from two sources: random measurement
errors, and systematic errors due to instrumental effects.  Our measurements
are technically challenging as we are measuring the diffuse emission,
requiring an absolute, not differential, measurement.  The IRS instrument
did not use a shutter, hence our measurements include the diffuse Milky Way
emission combined with zodical emission and instrumental backgrounds/artifacts.
This is in contrast to the usual differential measurements made with IRS
where off source observations are subtracted from on source observations
effectively removing the emissions in common (e.g., zodiacal and instrumental
emissions).  Figure~\ref{fig:image} shows a detector image for Hi~188.
The strong emission band at 11.3~\micron\ is clearly present in the long
wavelength (left) portion of the spectrum.  It is also obvious that there
is a non-uniform response in the spatial direction (perpendicular to the
dispersion) for the short wavelength (middle) portion of the spectrum.
This non-uniformity is most likely an instrumental effect, since it is
present in all of the spectra, regardless of their location with respect to
the Galaxy or the ecliptic.

These correlated errors are predominantly affect the SL1 spectra and depend
on detector location.  Because the sky is scanned by the observations, the
same wavelength or location on the sky can be observed at different detector
locations.  This gives rise to large, correlated variations which are mostly
detector based.  These large variations result in large off-diagonal elements
in the error covariance matrices and also create large contributions to the
random component of the diagonal elements (which are typically reported as
the mono-variate errors affecting the spectra).

To identify the correlated contributions, we performed principal axis
decomposition of the SL2 error covariance matrices. These showed that most
of the off-diagonal elements, and a large fraction of the diagonal elements,
resulted from a few well defined functions that are most likely related to
the cross-dispersion detector response seen in Figure~\ref{fig:image}.

Since it is much easier to manipulate and display mono-variate errors, and
because the signatures of the correlated errors are broad band and unlike
the lines we wish to analyze, we decided to correct the mono-variate errors
for the correlated contributions.  This was done by removing the
contributions of the two largest principal axes from the diagonal elements
of the error covariance matrices and then using these corrected diagonal
elements for error weighting and significance analyses.  This process was
only done for the SL2 data, as the off-diagonal elements of the SL1 data
were generally much smaller than the diagonal elements, indicating that
the magnitudes of the diagonal elements were not strongly affected by
correlated errors.

\begin{figure}
\begin{center}
  \includegraphics[width=0.9\linewidth]{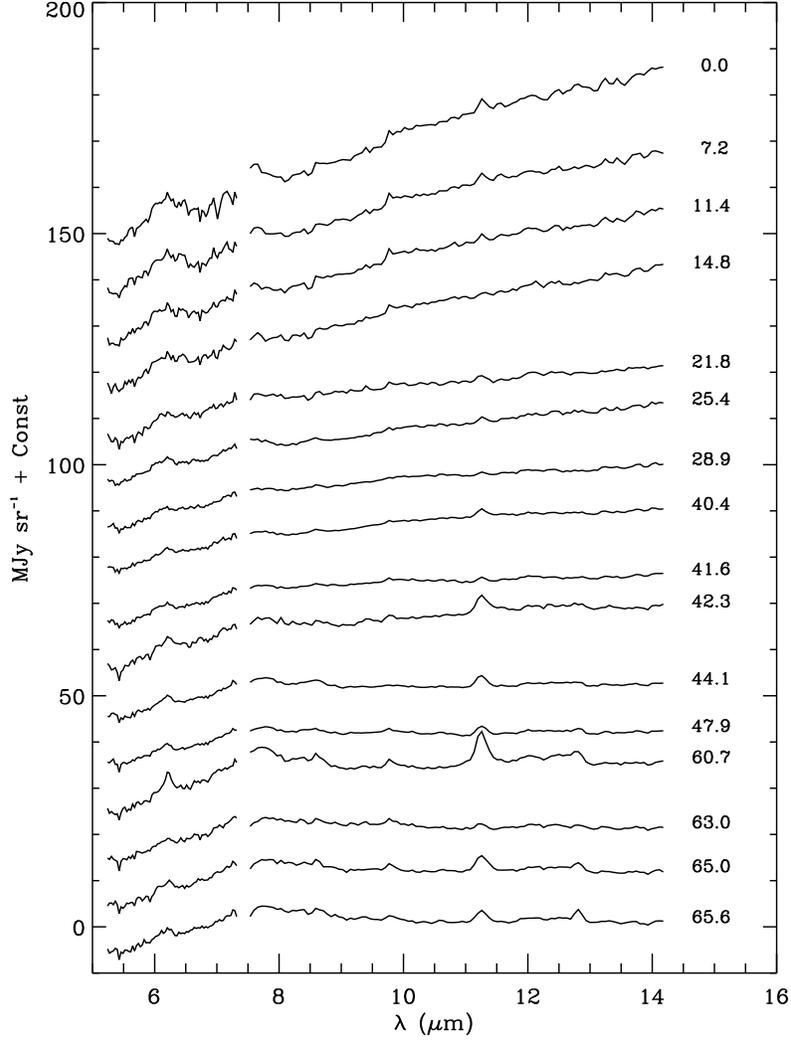}
\end{center}
\vspace{-0.8in}\caption{Spectra of the program stars in order of decreasing
absolute value of ecliptic latitude, $|\beta|$, listed to the right of each
spectrum.  The strong dependence of the continuum slope between 8 and 14.5
\micron\ on $|\beta |$ is clearly evident.  Also note that the spectral
feature  near 6.2 \micron\ is wider in spectra near the ecliptic plane.  This
appears  to result from the strengthening of a zodiacal light feature near
the 6.22 \micron\ interstellar PAH feature.}
\label{fig:ecliptic}
\end{figure}

\begin{figure}
\begin{center}
\includegraphics[width=0.5\linewidth]{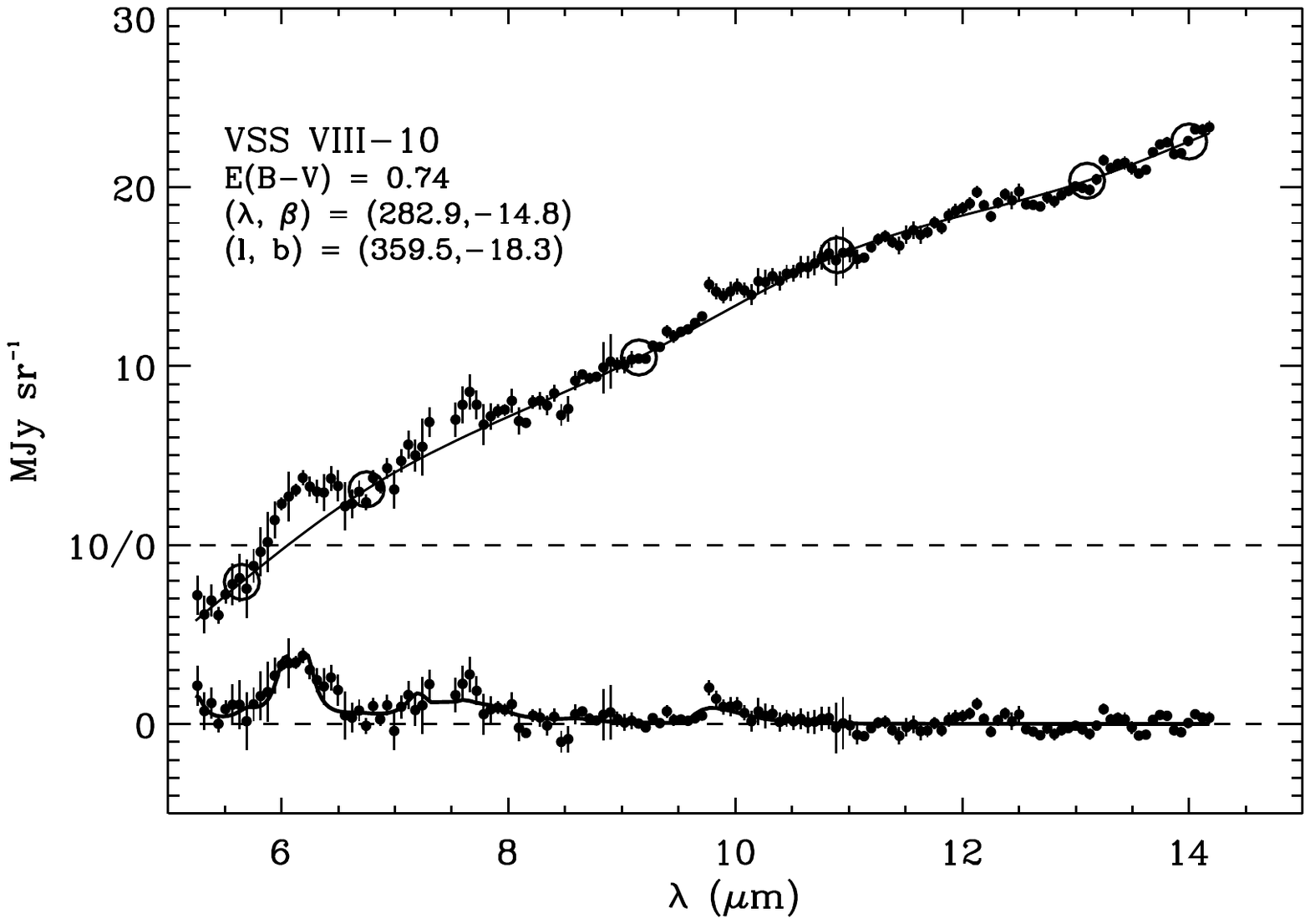}\hfill
\includegraphics[width=0.5\linewidth]{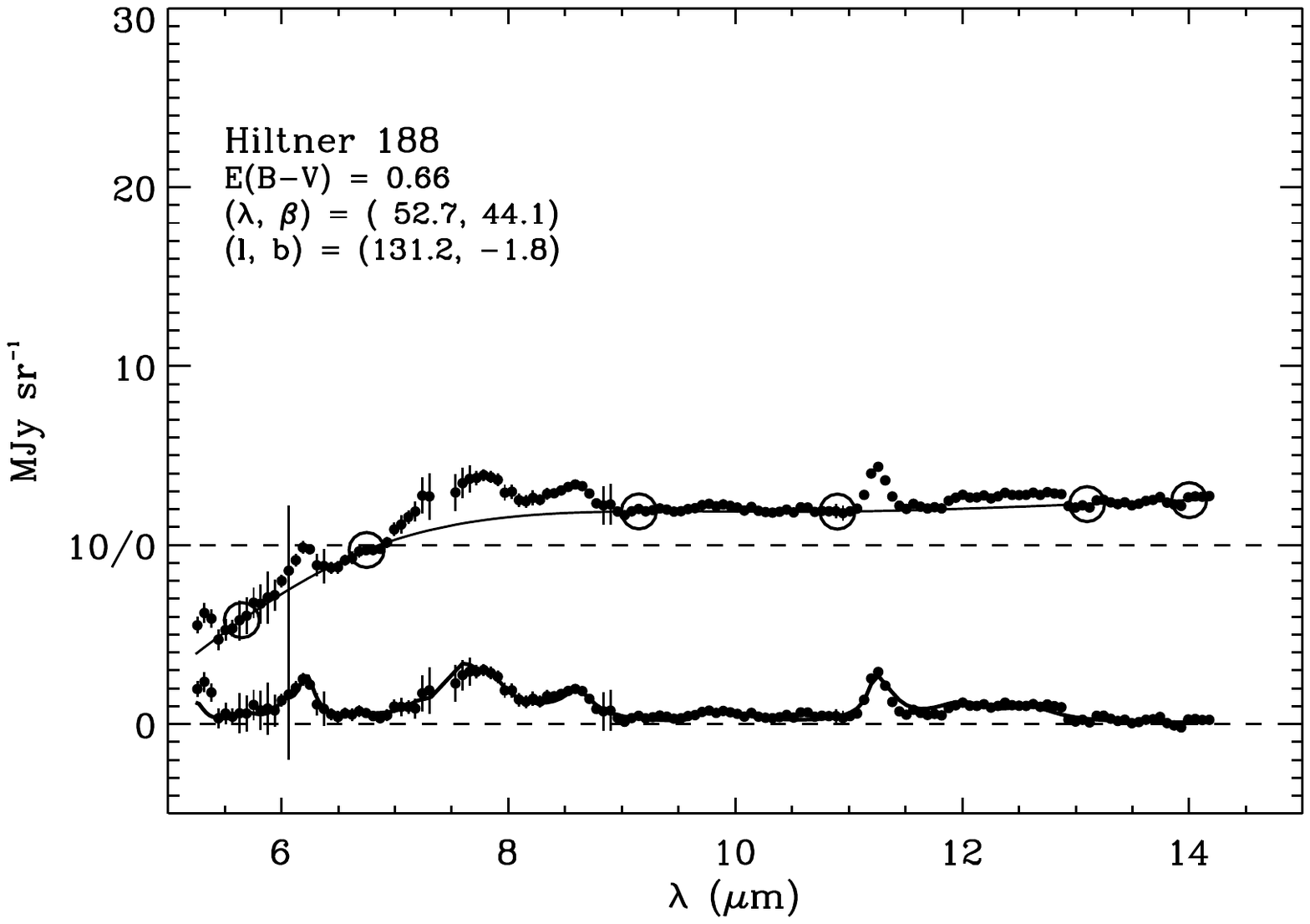}\vfill\vspace{-2.2in}
\vspace{0.4in}
\caption{Examples of spectra where, in each case, the upper plot is the
observed spectrum with the continuum background fit shown, and the lower
spectrum has the background subtracted. The circles are the locations of the
points used to determine the background continuum.}
\label{fig:examples}
\end{center}
\end{figure}

\section{Analysis \label{sec:analysis}}
In this section, we first describe how we model the zodiacal light
contribution to the observed spectra.  We then introduce a simple model
for the interstellar PAH emission lines.  Next, we consider additional
features that are present in the spectra but are not due to known PAH
features.  Finally, we define the model that we use to fit the observed
spectra and give the details of how the model is used to extract
quantitative information from the spectra.

\subsection{Zodiacal light continuum\label{sec:zodi}}
Figure~\ref{fig:ecliptic} shows the spectra toward the program stars,
arranged in order of the absolute magnitudes of their ecliptic latitude,
$|\beta|$, with those at the smallest latitudes at the top.  A few things are
apparent.  First, the overall slope of the continua increase with decreasing
$|\beta|$.  Second, the emission feature near 6.2 \micron\ is wider for
sight lines with small $|\beta|$.

To minimize the effects of the zodiacal light and to isolate and quantify
the interstellar features, we remove the continuous background.  This was
done by fitting a spline to 6 points which appear to be free of emission
features.  These points are at 5.65, 6.75, 9.15, 10.90, 13.1, and 14.0
\micron.  Figure~\ref{fig:examples} shows 2 examples of spectra with and
without the background subtracted.

\subsection{The PAH model\label{sec:pah}}
While a few of the PAH features are distinct, most of the main features in
the spectra are clearly blends.  Rather than attempting to fit every
possible emission feature in the spectra as an independent line, we adopt
the positions, widths and relative strengths of the Drude profiles used in
the \citet[][DL07 hereafter]{dl07}\defcitealias{dl07}{DL07}model as a
starting point.  However, instead of grouping the lines into ionized and
neutral PAH and with and without CH stretch, we simply use all of the lines
listed as ionized PAHs between 5 and 15 \micron, and break them into 4
segments, centered around the four strongest observed features, i.e., $5.0
\leq \lambda < 7$, $7 \leq \lambda < 8$, $8 \leq \lambda < 9$, and $9 \leq
\lambda \leq 15$.  We refer to these as the 6.2, 7.6, 8.5 and 11.3
\micron\ PAH features, respectively.

When comparing the observed spectra to the models, it became apparent that
a few of the lines had to be rescaled.  Consequently, we increased the
strength of the \citetalias{dl07} 5.25 \micron\ feature (which only has
its red wing in our wavelength range) and the strength of their 11.99
\micron\ feature by a factor of 2.5.

\subsection{Unidentified features}
Upon examining the continuum subtracted spectra, it became apparent that
they contained three features not included in the \citetalias{dl07}
model.  Consequently, we include three non-PAH features in our model.  These
are: 1) a feature we represented as a Drude profile with $\lambda_0 = 6.03$
\micron\ and $\gamma = 0.20$, $D(6.03, 0.29, \lambda)$, 2) a second feature
we approximate as a Drude profile with $\lambda_0 = 7.20$ \micron\ and
$\gamma = 0.25$, $D(7.20, 0.25,\lambda)$, and 3) an asymmetric feature
which we approximate as a $\chi^2$ distribution function with 5 degrees of
freedom and its abscissa, $u$, scaled and shifted so that $\lambda = a u +
\lambda_0$, where $a = 1/12$ and $\lambda_0 = 9.55$ \micron, $\chi^2_{5}
([\lambda-\lambda_0]/a)$.  These parameters place the peak of the
distribution near 9.8 \micron.  We refer to these three features as the
6.0, 7.2 and 9.8 \micron\ features and discuss their possible physical
origins in \S~\ref{sec:discussion}.

\begin{figure}
\begin{center}
\vspace{-.0in}\includegraphics[width=0.9\linewidth]{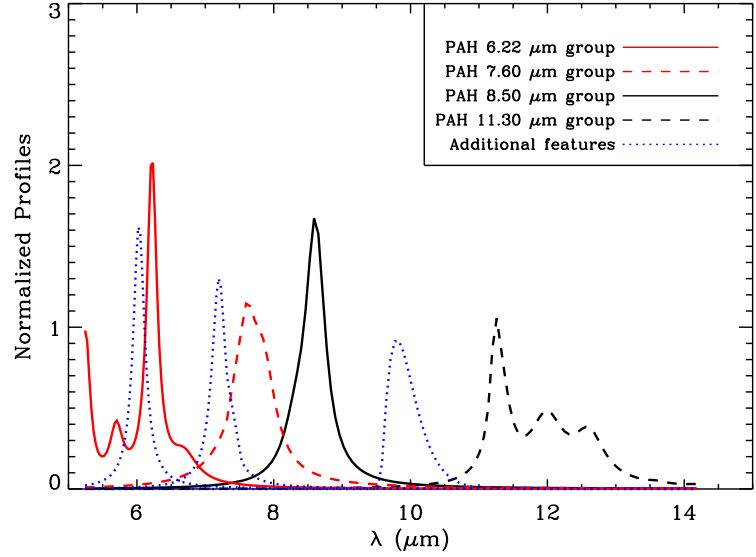}

\vspace{-3.3in}
\caption{The 7 fitting functions adopted for our analysis.
The four segments of the  \citetalias{dl07} PAH model we use are shown as
the solid red, dashed red, solid black and dashed black curves, which
represent the 6.22, 7.6, 8.5 and 11.3 \micron\ components, respectively.
The 3 dotted blue curves are the 3 additional features at 6.0, 7.2 and
9.8 \micron, which are described in the text.}
\label{fig:model_features}
\end{center}
\end{figure}

\subsection{The final model\label{sec:final}}
When fitting our model to the observations, we first fit a continuum to
each spectrum using the spline, described in \S\ref{sec:zodi}, and then
remove the continuum.  The resulting curve is then fit to the model described
above for seven free parameters, $c(\lambda_i)$, where $\lambda_i = 6.2, 7.6,
8.5, 11.3, 6.0, 7.2, 9.8$ \micron, i.e.,
\begin{equation}
f(\lambda) = \sum_{i=1}^4 c(\lambda_i)\Phi(\lambda)_i + \sum_{i=5}^6
             c(\lambda_i) D(\lambda_i, \gamma_i, \lambda) + c(\lambda_7)
             \chi^2_{5} (\lambda) -s(\lambda) \label{eq:model}
\end{equation}
where the $\Phi(\lambda)_i$ are the 4 segments of the \citetalias{dl07}
model (modified as described above), the two $D(\lambda_i, \gamma_i, \lambda)$
and $\chi^2_5(\lambda)$ are the two Drude profiles and $\chi^2$ distribution
described in the previous section and $s(\lambda)$ is the spline described
in \S~\ref{sec:zodi}, which is removed to place the model and the
observations into the same form. The fitting functions are shown in
Figure~\ref{fig:model_features}.  Because the profiles are all normalized,
the coefficients, $c(\lambda_i)$, have units of MJy~sr$^{-1}$.  The appendix
provides further details about the Drude profiles and explains how to
recover the contribution of each one form the $c(\lambda_i)$\ fit
parameters.  Using these functions, and the errors derived in
\S~\ref{sec:errors}, we performed non-linear least squares fits using the
Interactive Data Language (IDL) procedure MPFIT developed by
\cite{2009ASPC..411..251M}\footnote{The Markwardt IDL Library is available
at \url{http://purl.com/net/mpfit}}.  If we did not have to subtract the
spline in order to be consistent with the observations, we could simply
use a linear regression to fit the PAH and other features.  However,
because  removal of the spline from the fitting functions is required, it
becomes a non-linear least squares problem.

\begin{figure}
\begin{center}
  \includegraphics[width=1.0\linewidth]{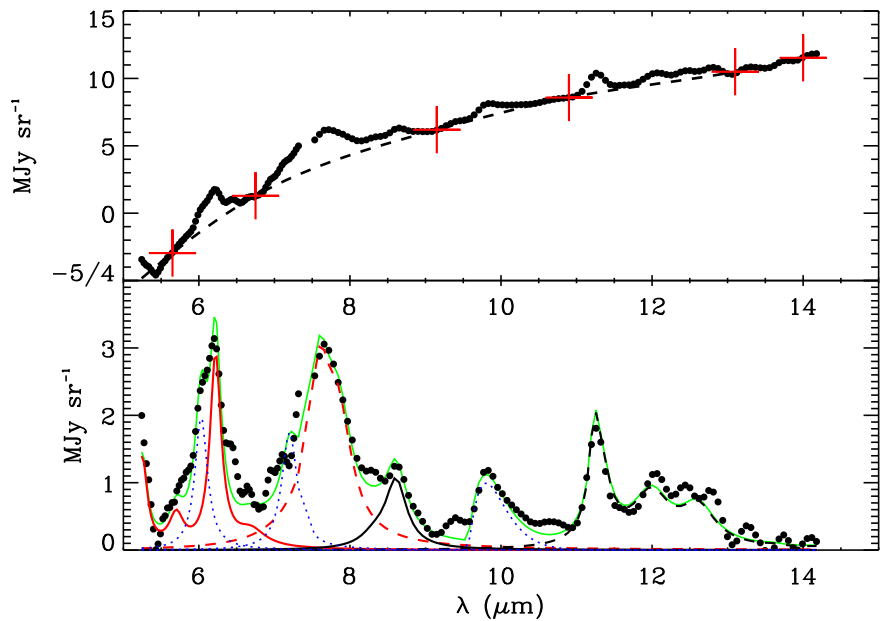}

\vspace{-3.5in}
\caption{Top: The points are the sample mean spectrum
(smoothed by 3 points) and the dashed curve is a spline fit to the 6 points
indicated by the large crosses.  Bottom: Green curve is an illustrative fit
to the mean spectrum with its continuum adjusted to agree with the model.
The solid, dashed red and black curves and dotted blue curves are the major
model features, with the same color code as Figure~\ref{fig:model_features}.
The relatively poor fit near 7 \micron\ results because this region is near
the poorly determined points at the edges of the SL1 and SL2 spectra.}
\label{fig:mean_fit}
\end{center}
\end{figure}

As an example of the fitting, Figure~\ref{fig:mean_fit} shows the fit to
the sample mean, and the scaled components of the fit.  Because zodiacal
light contamination precludes our ability to determine the continuum of
the observed spectra, the data have been adjusted to agree with the
continuum implied by the fit. Specifically, the spline fit to the observed
spectrum was removed and replaced with the spline fit to the model.  While
the overall fit to the strongest features is quite good, there are a few
regions where our model does not fit very well.  Specifically, these include
the regions of the minima near 6.5, 9.4 and 10.5 \micron, and the plateau
between 12 and 13 \micron.  This is not too surprising since the major
features are probably not perfect Drude profiles and several weaker features
have probably been omitted.  Nevertheless, the model appears to capture the
strongest features in the spectrum quite well, providing quantitative
measures of their strengths.

\begin{figure}
\begin{center}
\includegraphics[width=0.95\linewidth]{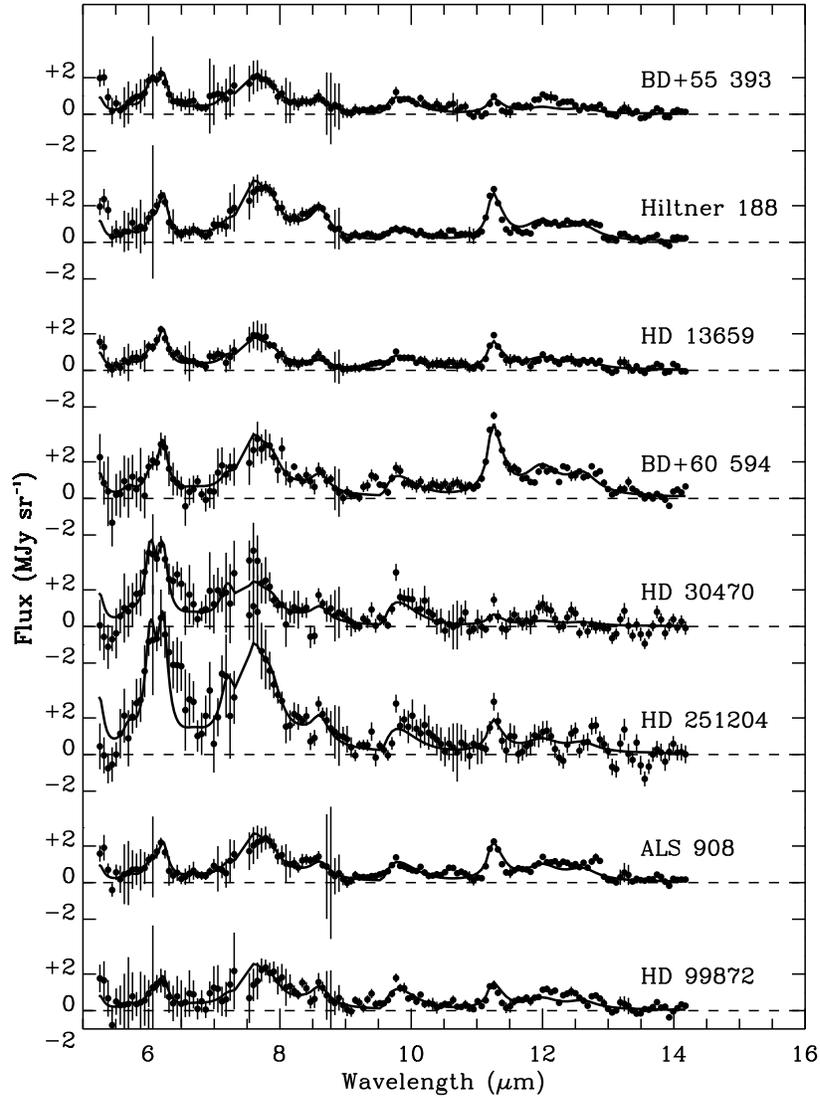}
\end{center}
\vspace{-1.0in}\caption{Stacked display of observed emission (points) and
one $\sigma$\ error bars, along with our model fits.  Each curve is labeled
by the star along its line of sight.}
\label{fig:stacks1}
\end{figure}
\newpage

\begin{figure}
\begin{center}
\includegraphics[width=0.95\linewidth]{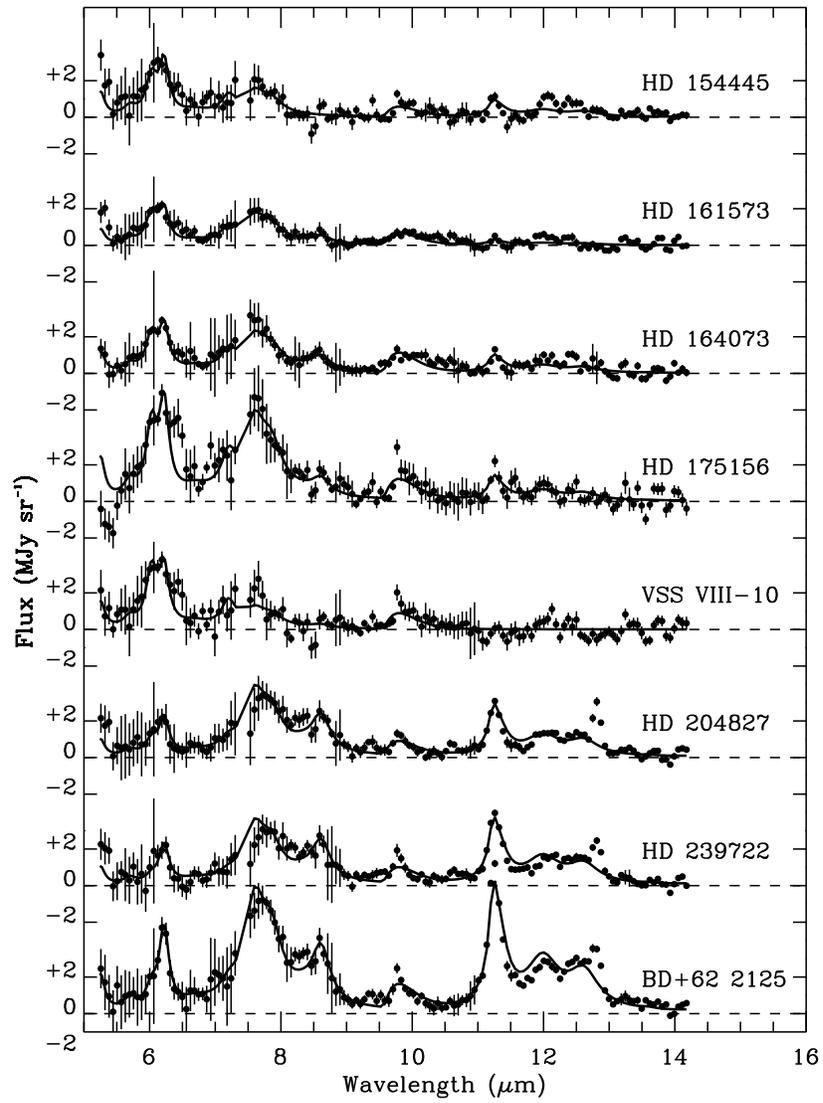}
\end{center}
\vspace{-1.0in}\caption{Same as Figure~\ref{fig:stacks1}.}\label{fig:stacks2}
\end{figure}

\clearpage

\section{Results\label{sec:results}}
We begin this section with a presentation of the fits to the spectra and a
description of a few of their properties.  Next, we investigate the
dependence of the model parameters on ecliptic latitude and then examine
the coefficients relations to each other.  Finally, we introduce a set
of extinction measurements for the program stars and examine correlations
between the extinction curve parameters and the parameters which describe
the \spitzer\ spectra.

\begin{table}[h]
\caption{Fit coefficients (in MJy sr$^{-1}$)}
\begin{center}
\begin{tabular}{lccccccc} \hline
Star        &  $c(6.2)$ & $c(7.6)$ & $c(8.5)$ & $c(11.3)$ &
$ c(9.8)$ & $c(6.0)$ & $c(7.2)$ \\ \hline
BD+55$^\circ$393    & $0.92 \pm 0.15$  & $1.75 \pm 0.10$  & $0.38 \pm 0.10$  & $0.81 \pm 0.05$  & $1.39 \pm 0.102$  & $0.46 \pm 0.32$  & $1.12 \pm 0.32$ \\
Hiltner 188  & $1.18 \pm 0.14$  & $2.83 \pm 0.08$  & $0.96 \pm 0.08$  & $2.51 \pm 0.04$  & $1.00 \pm 0.085$  & $0.25 \pm 0.28$  & $1.04 \pm 0.28$ \\
HD 13659     & $0.98 \pm 0.10$  & $1.49 \pm 0.07$  & $0.25 \pm 0.07$  & $1.49 \pm 0.07$  & $1.19 \pm 0.104$  & $0.27 \pm 0.21$  & $1.12 \pm 0.21$ \\
BD+60$^\circ$594    & $1.40 \pm 0.23$  & $2.93 \pm 0.11$  & $0.64 \pm 0.11$  & $3.83 \pm 0.10$  & $1.74 \pm 0.190$  & $0.34 \pm 0.32$  & $0.49 \pm 0.32$ \\
HD 30470     & $1.78 \pm 0.23$  & $1.93 \pm 0.13$  & $0.53 \pm 0.13$  & $0.60 \pm 0.18$  & $2.01 \pm 0.276$  & $1.20 \pm 0.68$  & $1.77 \pm 0.68$ \\
HD 251204    & $3.07 \pm 0.46$  & $4.96 \pm 0.14$  & $0.96 \pm 0.14$  & $1.75 \pm 0.25$  & $1.95 \pm 0.335$  & $1.82 \pm 0.69$  & $1.62 \pm 0.69$ \\
ALS 908      & $0.95 \pm 0.14$  & $2.26 \pm 0.08$  & $0.55 \pm 0.08$  & $2.09 \pm 0.04$  & $1.65 \pm 0.095$  & $0.27 \pm 0.33$  & $1.39 \pm 0.33$ \\
HD 99872     & $0.81 \pm 0.28$  & $2.14 \pm 0.14$  & $0.67 \pm 0.14$  & $1.45 \pm 0.06$  & $1.75 \pm 0.140$  & $0.21 \pm 0.45$  & $1.53 \pm 0.45$ \\
HD 154445    & $1.42 \pm 0.27$  & $1.30 \pm 0.00$  & $0.00 \pm 0.00$  & $0.90 \pm 0.11$  & $0.86 \pm 0.150$  & $0.62 \pm 0.44$  & $1.10 \pm 0.44$ \\
HD 161573    & $0.89 \pm 0.13$  & $1.50 \pm 0.10$  & $0.24 \pm 0.10$  & $0.28 \pm 0.07$  & $1.08 \pm 0.113$  & $0.50 \pm 0.30$  & $0.98 \pm 0.30$ \\
HD 164073    & $1.18 \pm 0.10$  & $1.92 \pm 0.12$  & $0.50 \pm 0.12$  & $0.97 \pm 0.08$  & $1.72 \pm 0.123$  & $0.60 \pm 0.31$  & $1.12 \pm 0.31$ \\
HD 175156    & $2.45 \pm 0.23$  & $4.13 \pm 0.13$  & $0.58 \pm 0.13$  & $1.31 \pm 0.18$  & $1.76 \pm 0.261$  & $1.16 \pm 0.46$  & $2.21 \pm 0.46$ \\
VSS VIII-10  & $1.56 \pm 0.21$  & $1.01 \pm 0.12$  & $0.11 \pm 0.12$  & $0.00 \pm 0.00$  & $1.40 \pm 0.227$  & $0.94 \pm 0.42$  & $0.98 \pm 0.42$ \\
HD 204827    & $1.00 \pm 0.28$  & $3.34 \pm 0.14$  & $1.27 \pm 0.14$  & $2.75 \pm 0.07$  & $1.19 \pm 0.149$  & $0.24 \pm 0.42$  & $0.91 \pm 0.42$ \\
HD 239722    & $1.05 \pm 0.21$  & $3.09 \pm 0.16$  & $1.36 \pm 0.16$  & $3.51 \pm 0.08$  & $1.34 \pm 0.152$  & $0.11 \pm 0.40$  & $1.39 \pm 0.40$ \\
BD+62$^\circ$2125   & $2.16 \pm 0.25$  & $5.85 \pm 0.17$  & $1.95 \pm 0.17$  & $6.80 \pm 0.10$  & $2.12 \pm 0.154$  & $0.33 \pm 0.70$  & $1.27 \pm 0.70$ \\ \hline
\end{tabular}
\end{center} \label{tab:fitparams}
\end{table}

Table~\ref{tab:fitparams} lists the parameters that result from fits to
the data using eq.~(\ref{eq:model}).  Figures~\ref{fig:stacks1} and
\ref{fig:stacks2} display the fits to the individual lines of sight
and, as in Fig.~\ref{fig:mean_fit}, we display the fits using the model
continuum derived for each one.  A few features are worth emphasizing.
Generally, the strength of the features scale with \ebv, but there are
clear exceptions.  For example, BD$+60^\circ 594$\ has a slightly smaller
color excess than VSS VIII-10, but the PAH features are much stronger in
BD$+60^\circ 594$.  In contrast, the features in the spectra toward HD~13659
and HD 161573 have similar strengths, even though their $E(B-V)$ differ by a
factor of 4.  Generally, the strengths of the PAH features appear to vary
together, although there are a few exceptions.  Specifically, for most sight
lines, the 7.6 \micron\ feature is only slightly stronger than the 11.3
\micron\ feature, but toward HD~239722 and HD~175156, the peak of the 7.6
\micron\ feature is far stronger.  However, we shall argue below that this
may actually result from contamination of the 7.6 \micron\ feature in
HD~251204 and HD~175156 by zodiacal emission.  Finally, we point out that
the [Ne {\sc ii}] 12.81 \micron\ emission line is present in the sight
lines toward HD~204827 and HD~239722, both in the open cluster Tr~37, and
probably present in the line of sight to BD$+62^\circ$ 2125, a member of
Cep~OB3.

\begin{figure}
\begin{center}
\includegraphics[width=0.7\linewidth]{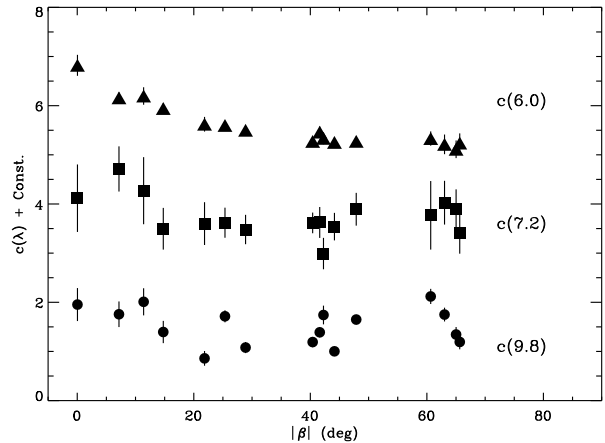}
\end{center}
\vspace{-2.7in}\caption{Coefficients of the unidentified features plotted
against the absolute value of the ecliptic latitude of their line of sight.}
\label{fig:zodi_features}
\end{figure}

We begin our search for correlations by examining the relations of the
unidentified features to the absolute value of ecliptic latitude, $|\beta|$.
Figure~\ref{fig:zodi_features} presents these coefficients plotted against
$|\beta|$\ for each line of sight.  There is no obvious correlation for the
9.8 \micron\ feature, and the 7.2 \micron\ feature may be weakly correlated
with $|\beta|$.  However, the 6.0 \micron\ line is strongly correlated with
$|\beta |$.

\begin{figure}
\begin{center}
\includegraphics[width=1.0\linewidth,angle=180]{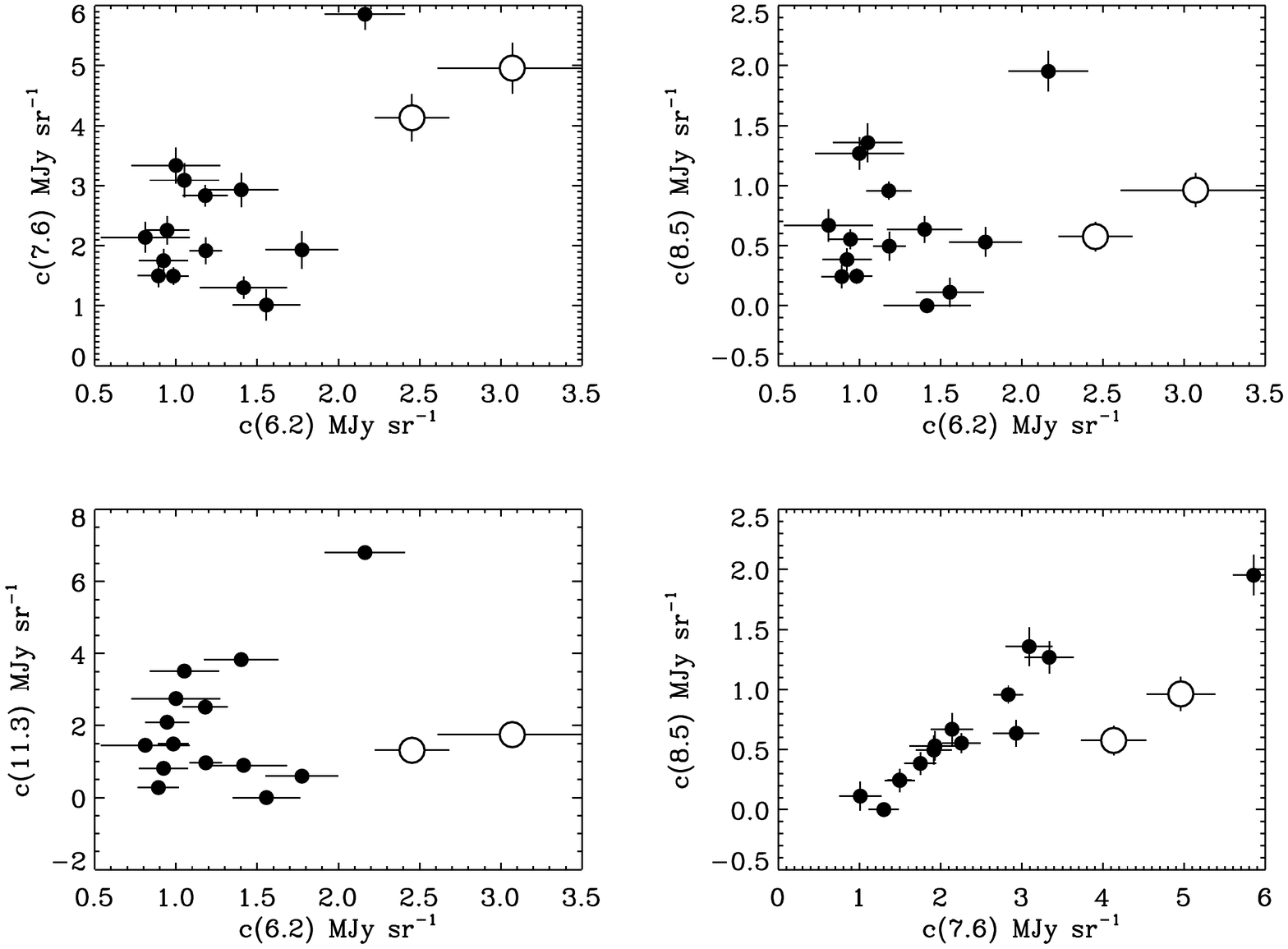}\vfill
\vspace{-0.8in}
\includegraphics[width=1.0\linewidth,angle=180]{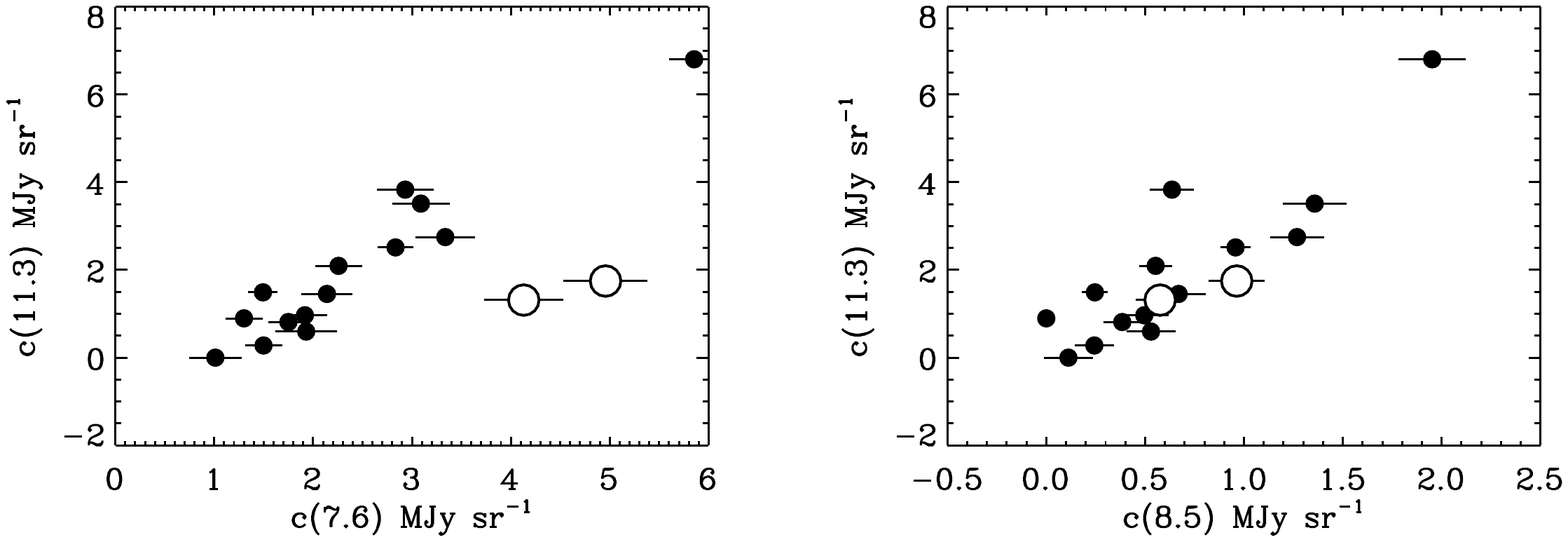}
\end{center}
\vspace{-2.3in}\caption{The strengths of the four PAH features plotted
against each other.  One $\sigma$\ error bars are also shown.
Open circles are for stars with $|\beta| < 10^\circ$.}
\label{fig:pah-pah}
\end{figure}

\begin{figure}
\begin{center}
\includegraphics[width=0.5\linewidth,angle=180]{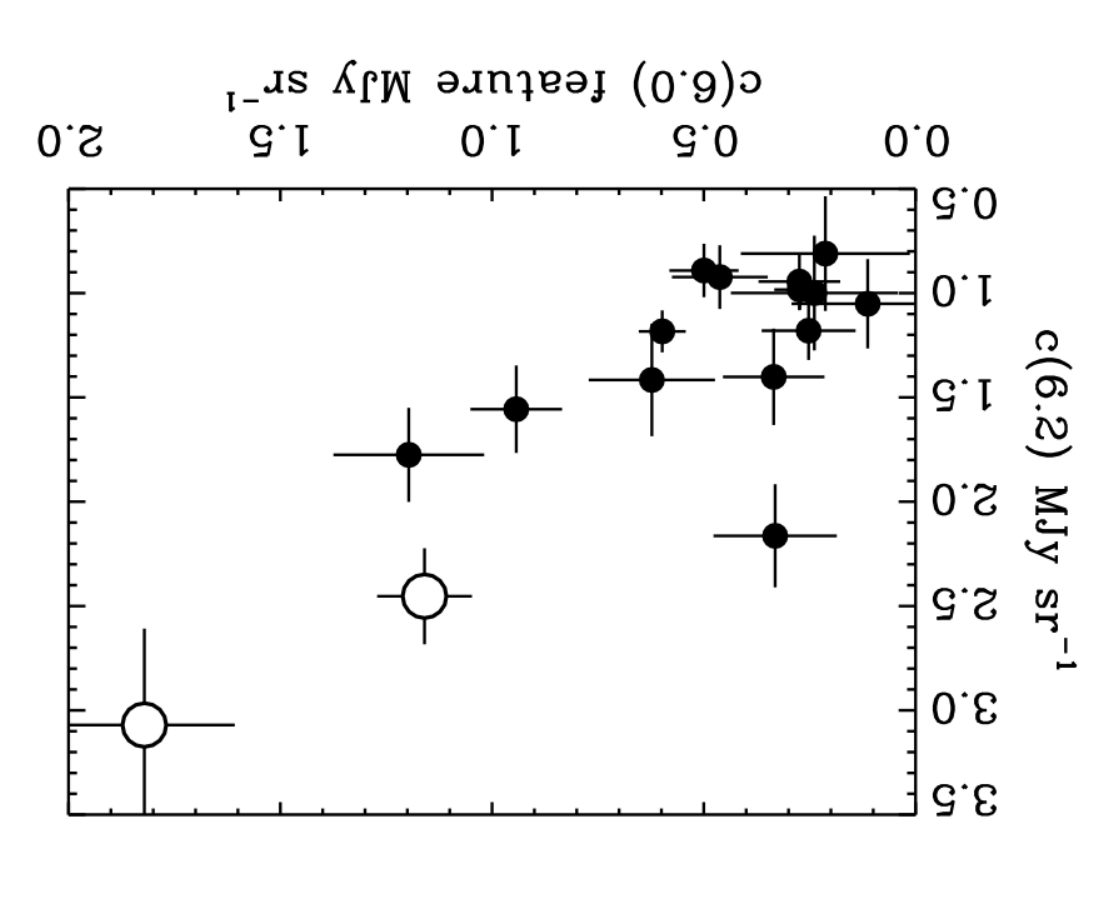}
\end{center}
\vspace{-0.2in}
\caption{Correlation between the 6.2 \micron\ PAH feature and the 6.0
\micron\ unidentified feature.  Open circles are for stars with $|\beta| <
10^\circ$.}
\label{fig:c6_c6}
\end{figure}

Figure \ref{fig:pah-pah} shows the strengths of the four PAH features
plotted against each other.  Clearly, the 6.2 \micron\ PAH feature is not
correlated with the others, but the 7.6, 8.5 and 11.3 \micron\ PAH features
are all strongly related to one another.  On the other hand, Figure
\ref{fig:c6_c6} shows that the 6.2 \micron\ PAH feature is correlated with
6.0 \micron\ feature, which has been shown to be related to the zodiacal
light \cite{2003Icar..164..384R}.  It is likely that the reason the 6.2
\micron\ feature is not correlated with the other PAH features is that the
values we derive for its strength are entangled with those for the 6.0
\micron\ feature.

The two points shown as open circles in fig. \ref{fig:pah-pah} deviate
from the others for plots involving $c(7.6)$.  These are for HD 251204
and and HD~175146.  They have the smallest $|\beta|$\ values in our
sample, $0.0^\circ$\ and $7.2^\circ$, respectively (see,
Table~\ref{tab:sample}).  This makes their spectra the most susceptible
to zodiacal light contamination.  Thus, we suspect that zodiacal light
affects not only $c(6.2)$, but probably $c(7.6)$ as well for these two
stars.

We now turn to the extinction parameters.  Some of these are the
coefficients from eq.~(2) in \citet{fm07}, which we denote as $a_i$.
Specifically, we use the following parameters: \ebv\ (the color excess),
$A(V) = E(B-V) \times R(V)$ (the total extinction in the $V$ band), $a_2
\times E(B-V)$\ (a measure of the slope of the UV extinction), $a_4 \times
E(B-V)$\ (a measure of the curvature of the far-UV extinction), and \abump\
(the bump area), which is given by

\begin{equation}
A(2175) = \frac{\pi a_3}{2 \gamma} E(B-V)
\end{equation}
The observed values of these parameters for the program stars are listed in
Table~\ref{tab:properties}

\begin{table}[h]
\caption{Extinction Properties}
\begin{center}
\begin{tabular}{lcccc}
Star    & $E(B-V)$ & \abump & $A(V)$  & $a_2 E(B-V)$ \\ \hline
BD+55$^\circ$393    & 0.26  & 1.172  & 0.738 &  0.179\\
Hiltner 188  & 0.66  & 3.487  & 1.888 &  0.667\\
HD 13659     & 0.80  & 2.631  & 1.984 &  0.920\\
BD+60$^\circ$594    & 0.65  & 3.579  & 1.729 &  0.507\\
HD 30470     & 0.34  & 1.879  & 1.108 &  0.180\\
HD 251204    & 0.76  & 3.523  & 2.333 &  0.555\\
ALS 908      & 0.65  & 3.241  & 1.943 &  0.455\\
HD 99872     & 0.32  & 2.267  & 0.957 &  0.067\\
HD 154445    & 0.40  & 2.734  & 1.180 &  0.140\\
HD 161573    & 0.20  & 1.070  & 0.620 &  0.072\\
HD 164073    & 0.21  & 0.504  & 1.121 &  0.109\\
HD 175156    & 0.35  & 1.615  & 1.095 &  0.438\\
VSS VIII-10  & 0.74  & 1.427  & 3.093 &  0.673\\
HD 204827    & 1.09  & 4.834  & 2.681 &  1.330\\
HD 239722    & 0.88  & 5.014  & 2.297 &  0.766\\
BD+62$^\circ$2125   & 0.91  & 4.871  & 2.466 &  0.692\\  \hline
\end{tabular}
\end{center}\label{tab:properties}
\end{table}

When comparing dust absorption and emission, it is important to bare in
mind that the absorption depends simply on the column density of absorbers
along the line of sight.  In contrast, the emission also depends on the
ambient radiation field along the line of sight and the uniformity of the
emission over the solid angle subtended by the observing aperture.

Our method of selecting targets maximizes the chance that they are beyond
the vast majority of dust along the line of sight and do not interact
with it.  Nevertheless, the sight line may encounter complex regions
where the radiation field is very different from the mean and the spatial
distribution of the dust is highly variable.

\begin{figure}
\begin{center}
\includegraphics[width=0.5\linewidth]{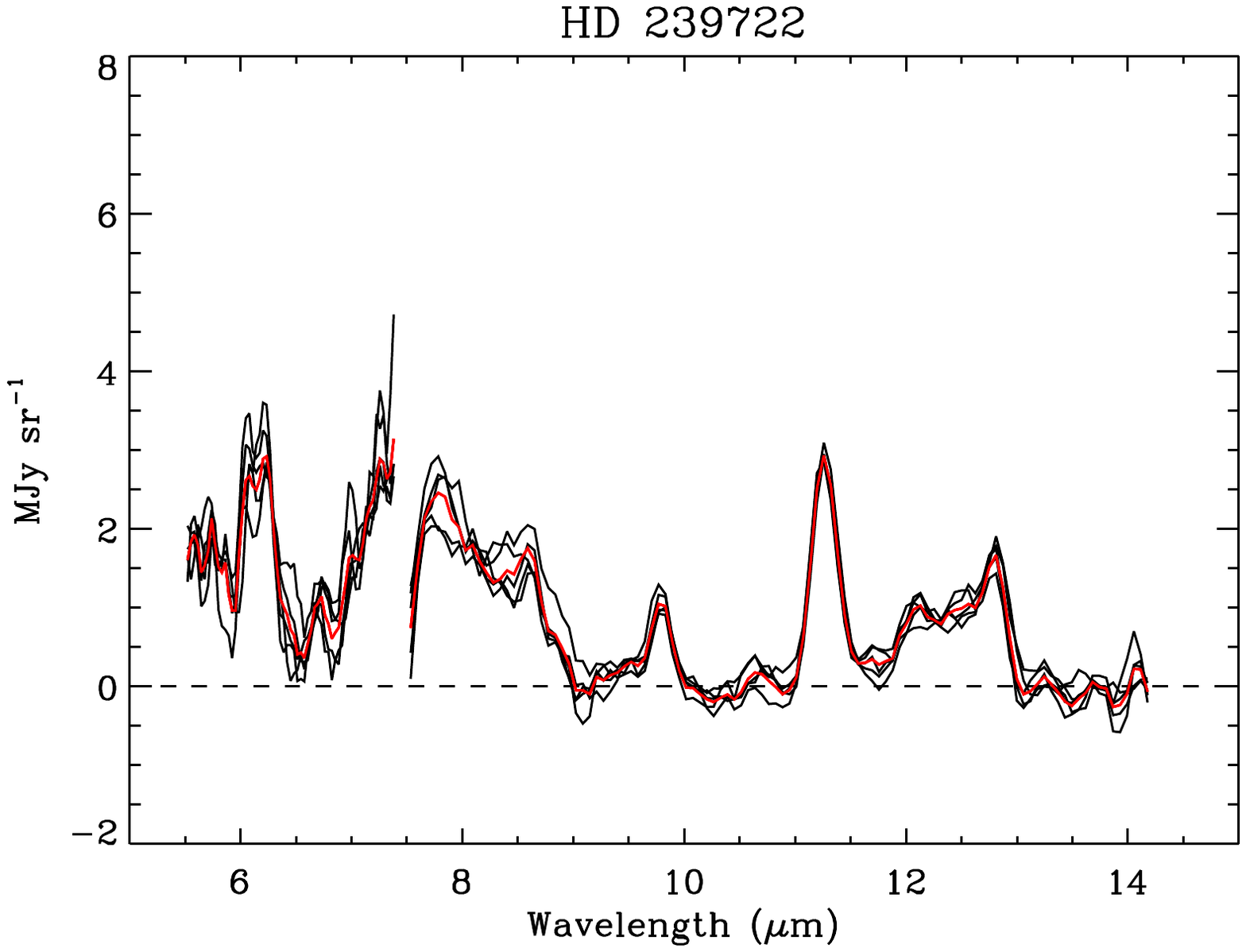}\hfill
\includegraphics[width=0.5\linewidth]{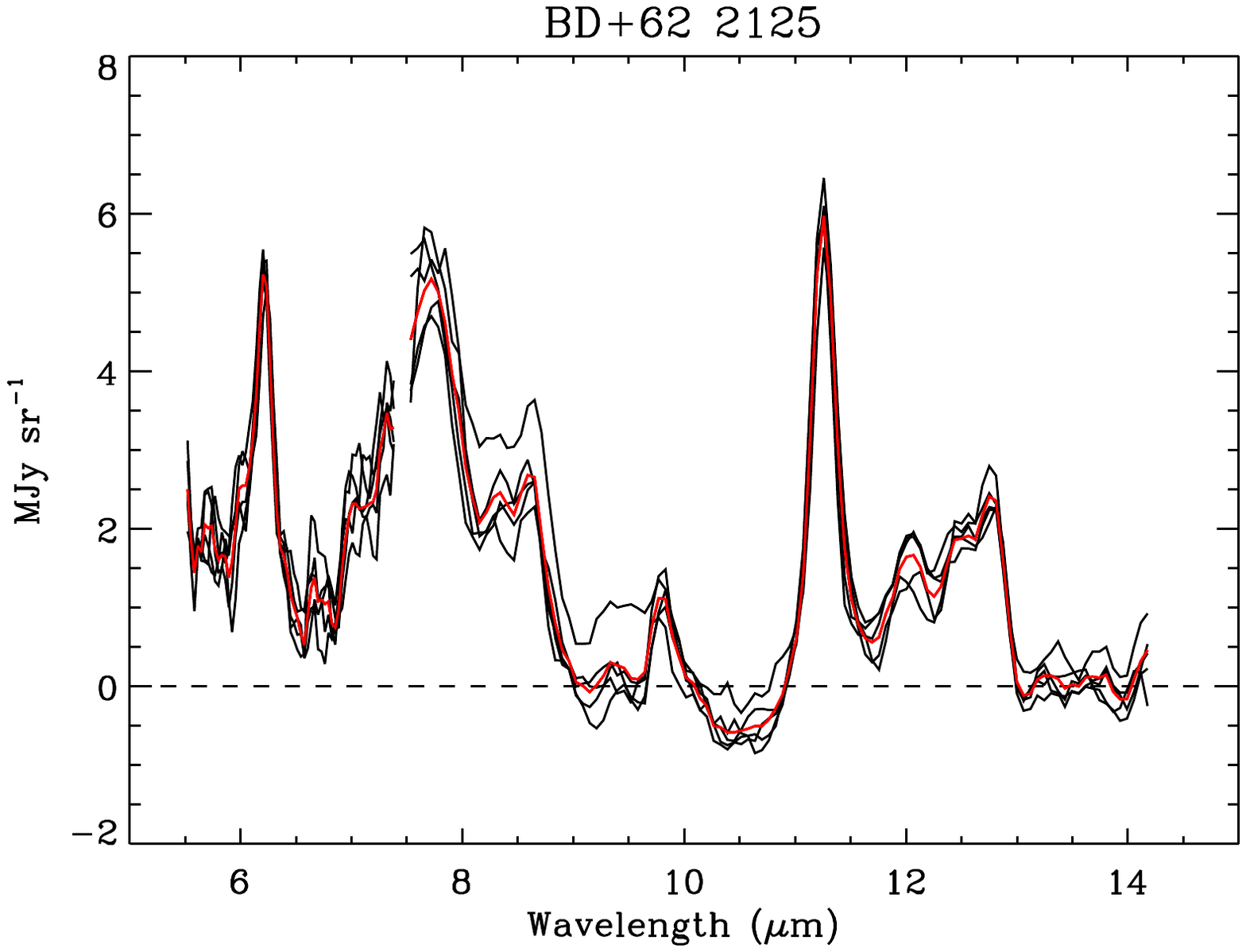}
\end{center}
\vspace{-1.7in}\caption{
SL1 and SL2 spectra for all of the different aperture positions (black) and
the mean (red) for HD~239722 (left) and BD$+62^\circ$ 2125.  All spectra
were smoothed by 3 points.  The scatter in the HD~239722 spectra is typical
of the stars in our sample.  In contrast, BD$+62^\circ$ 2125 is the only
star in the sample with a spectrum that appears systematically different
from the others.
}
\label{fig:subspec}
\end{figure}

The uniformity of the radiation field over the solid angle of the aperture
can be evaluated by examining the variance in the 6 slit positions observed
for each star (see figure~\ref{fig:wavecut}).  The combined SL1 and SL2
observations sample a region centered on the star that is roughly
$215^{\prime \prime}$ in $x$ and $45^{\prime \prime}$ in $y$.  Figure
\ref{fig:subspec} shows two sets of these observations; one is for a star
whose whose deviations are typical and one for BD$+62^\circ$ 2125, which has
largest systematic deviations in the sample.  It is clear that one spectrum
has considerably more emission over the region $ 8 \leq \lambda \leq 10$
\micron.  Nevertheless, even in this extreme case, the overall strength of
the major PAH features are not strongly affected.

\begin{figure}
\begin{center}
\includegraphics[width=1.0\linewidth,angle=180]{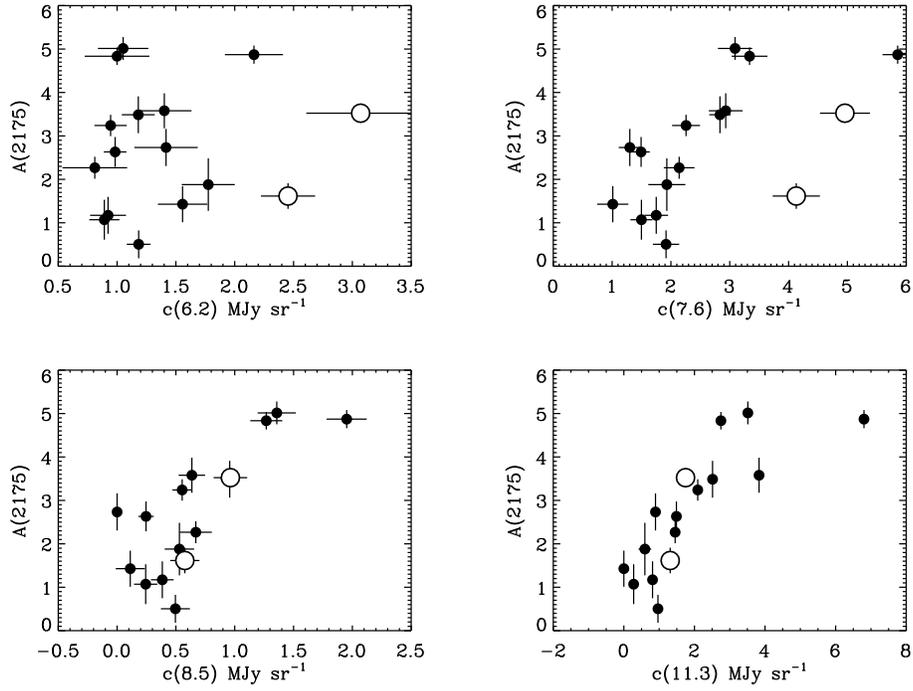}\vfill
\vspace{-0.8in}
\end{center}
\vspace{.4in}\caption{The area of the \bump\ bump plotted against the
strengths of the four PAH features, with $1 \sigma$\ error bars.  Open
circles are for stars with $|\beta| < 10^\circ$.}
\label{fig:2175}
\end{figure}

Figure~\ref{fig:2175} shows how the four PAH features are related to
\abump.  We see that \abump\ is strongly correlated to the 8.5 and 11.3
\micron\ features, somewhat correlated with the 7.6 \micron\ feature, but
uncorrelated to the 6.2 \micron\ feature.  However, we have already shown
that the 6.2 \micron\ data are contaminated by zodiacal emission, so this
last result is not surprising.  Further, when the two stars with $|\beta |
< 10^\circ$\ (again shown as open circles) are omitted, the relationship
between $c(6.7)$ and \abump\ is also quite strong.

\begin{figure}
\begin{center}
\includegraphics[width=1.0\linewidth,angle=180]{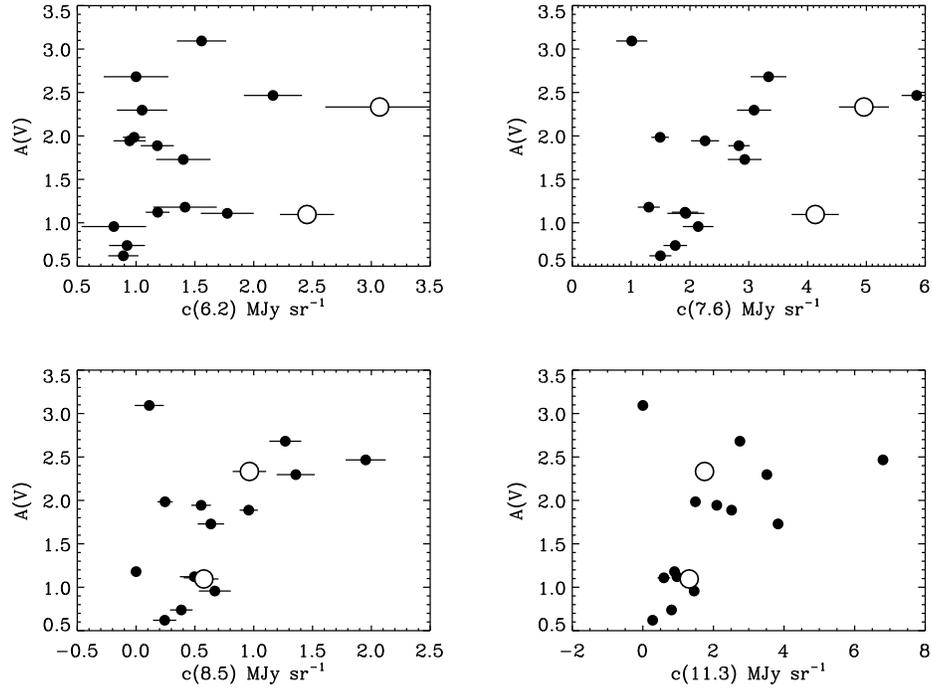}\vfill
\vspace{-0.8in}
\end{center}
\vspace{.4in}\caption{$A(V)$\  plotted against the strengths of the four PAH
features, with $1 \sigma$\ error bars.  Open circles are for stars with
$|\beta| < 10^\circ$.
}
\label{fig:av}
\end{figure}

Figure~\ref{fig:av} presents the relations between the total optical
extinction, $A(V)$, and the PAH parameters.  In this case, no strong
correlations are present.  Together with the previous plot, these results
indicate that the PAH emission is strongly correlated with whatever causes
the \bump\ absorption, but not with the overall line of sight extinction.

We also examined correlations between the PAH emission features and the
dust temperature, as determined by \citet{1998ApJ...500..525S}.  However,
no significant relations were found.

\clearpage
\section{Discussion \label{sec:discussion}}

In this section, we summarize and discuss our results.  We begin with the
results for the features not included in the \citetalias{dl07} model.

\begin{itemize}

\item The 6.0 \micron\ feature is strongly related to ecliptic latitude,
implying that it is related to the zodiacal light.  Furthermore, it seems to
be present in some of the zodiacal light spectra presented in
\cite{2003Icar..164..384R}.   It is also correlated with the 6.2 \micron\
PAH feature.  This suggests that the 6.2 \micron\ and 6.0 \micron\
measurements are entangled, and probably accounts for lack of correlation
between the 6.2 \micron\ feature and the other PAH features.

\item The 7.2 \micron\ feature is weakly correlated to $|\beta|$ and
unrelated to anything else.  We suspect that it is primarily an adjustment
which accounts for any residual discontinuity between the SL1 and SL2
caused by the "dark settle" effect described in \S~\ref{sec:data}.
Errors in this discontinuity may be larger when the zodiacal emission is
stronger, which could account for the weak correlation with $|\beta |$.

\item The 9.8 \micron\ emission feature is near the 9.7 \micron\ silicate
absorption feature, suggesting a possible relation.  However, it also
appears to be present in the zodiacal light spectra presented in
\cite{2003Icar..164..384R}, but it is not significantly correlated with
ecliptic latitude, so its origin is uncertain.

\item The [Ne {\sc ii}] 12.81 \micron\ emission line is present in three
of the sight lines in Figure~\ref{fig:stacks2}.  It is quite strong in
HD~204827 and HD~2397223 (both members of Tr 37) and a bit weaker toward
BD$+62^\circ 2125$, a member of Cep~OB3.  Both of these regions are sites
of recent and ongoing star formation.
\end{itemize}

We now summarize our results for the PAH features.
\begin{itemize}
    \item The 8.5 and 11.3 \micron\ PAH features at are strongly related
    to one another.

    \item 7.8 \micron\ PAH feature is also strongly correlated with the
    8.5 and 11.3 \micron\ features with the exception of two stars.
    These have the smallest $|\beta |$\ in the sample and we suspect that
    this is due to zodical light contamination (possibly from the 7.2
    \micron\ feature).

    \item The 6.2 \micron\ PAH feature does not correlate well with the
    other PAH features.  This is likely because its  values are entangled
    with the nearby 6.0 \micron\ unidentified, possible zodiacal, feature.

    \item The bump area, \abump, is strongly correlated with the
    8.5 and 11.3 \micron\ features, less strongly correlated with
    7.6 \micron\ feature, and uncorrelated to the 6.2 \micron\
    feature, mimicking how the PAH features are related to each
    other.  The one star whose data deviate somewhat from this relation
    is BD$+62^\circ 2125$, whose line of sight (which passes through the
    star forming region Cep~OB3) seems to have more PAH intensity than
    expected for its \abump.

    \item The total extinction at optical wavelengths, $A(V)$, is not
    correlated to the strengths of the PAH parameters.  This argues that the
    relation between \abump\ and the strengths of the PAH features is due to
    a similarity in the kind of dust along the line of sight and not just
    the amount.
\end{itemize}

Our first main finding is that the 7.6, 8.5 and 11.3 \micron\ PAH features
are strongly correlated.  The correlations among the PAH lines indicate
that the intensities of these features respond similarly.  Recall that
there are cases where the amount of interstellar dust, as measured by
$E(B-V)$, is quite large and the emission features are weak, and vice
versa.  Thus, whatever produces these features, affects all three in the
same way.  Thus, it appears that the ratios of the amplitudes of the PAH
features are roughly constant for our lines of sight through the relatively
diffuse interstellar medium.  In fact, simple regressions of $c(7.6)$ and
$c(8.5)$ on $c(11.3)$, with the intercepts forced to 0, give slopes
(ratios) of $c(7.6)/c(11.3) = 0.96$ (omitting HD~251204 and HD~175156) and
$c(8.5)/c(11.3) = 0.31$ (including all stars).  Consequently, for $c(7.6)$,
$c(8.5)$ and $c(11.3)$, the line of sight environment affects their
brightness, but not their ratios.

The second main finding is that \abump\ is strongly correlated with the
strength of the PAH emission lines.  In some ways, this strong correlation
is surprising.  This is because \abump\ is a direct measure of the total
column density of \bump\ absorbers along the line of sight, while the PAH
emission depends not only on the number of emitters, but also on the local
radiation field at each point along the line of sight.  Thus, even if the
carriers of the \bump\ feature and the PAH emitters were identical, this
would not guarantee a correlation between \abump\ and the strength of the
PAH emission.  For such a relation to occur, either the diffuse radiation
field along each line of sight must be roughly constant, or the conditions
that create the \bump\ absorbers and the PAH emitters are {\em both}
directly related the ambient radiation field,
perhaps through a two step process of the sort described by
\cite{2006ApJ...636..303W} in a much different context.

\acknowledgements
This work is based on observations made with the Spitzer Space Telescope,
which was operated by the Jet Propulsion Laboratory, California Institute of
Technology under a contract with NASA. Support for this work was provided by
NASA through an award issued by JPL/Caltech.  We thank Karin Sandstrom for
providing us with the "dark settle" correction IDL code.  We also thank
an anonymous referee for a thorough reading of the manuscript and useful
suggestions.  This research has made use of the NASA/IPAC Infrared Science
Archive, which is operated by the Jet Propulsion Laboratory, California
Institute of Technology, under contract with the National Aeronautics and
Space Administration.

\clearpage
\appendix
This appendix gives a detailed explanation of the Drude profiles used in our
PAH model and how to disentangle the contributions of each specific
component.

A Drude profile, $D(\lambda_i, \gamma_i, \lambda)_0$\ is defined by its
central wavelength, $\lambda_i$, and $\gamma_i$, which determines its width.
The wavelength dependence of the profile is given by

\begin{equation}
D(\lambda_i, \gamma_i, \lambda)_0 = \frac{\gamma_i \lambda_i}{(\lambda/\lambda_i
  -\lambda_i/\lambda)^2 +\gamma_i^2}
\end{equation}

In this paper, we use normalized Drude profiles,

\begin{equation}
D(\lambda_i, \gamma_i, \lambda) = \frac{D(\lambda_i, \gamma_i)_0}{ \int_0^\infty
                         D(\lambda_i, \gamma_i)_0}
\end{equation}

The PAH fitting functions, $\Phi(\lambda)_i$, used in equation~(\ref{eq:model})
are composed of several normalized Drudes, multiplied by coefficients,
$b_{i,j}$, which are proportional to their relative contributions in the
\citetalias{dl07} models.  Since the $\Phi(\lambda)_i$ are normalized to
unity, we have $\sum_{j=1}^N b_{i,j}= 1$, for all $i$, and

\begin{equation}
\Phi(\lambda)_i = \sum_{j=1}^N b_{i,j} D(\lambda_{i,j}, \gamma_{i,j}, \lambda)
\end{equation}
The values of $\lambda_{i,j}$, $\gamma_{i,j}$ and $b_{i,j}$ are given in
Table~\ref{tab:fractions}.
As a result, it is possible to use the $c(\lambda_i)$\ fit parameters given
in equation~(\ref{eq:model}) together with the $b_{i,j}$\ to determine the
contribution of each Drude to the fit, i.e.,  $c(\lambda_i) \; b_{i,j} \;
D(\lambda_{i,j}, \lambda_{i,j}, \lambda)$ gives the contribution of the
$j^{th}$ Drude to $\Phi(\lambda)_i$.

\begin{table}[h]
\caption{$\Phi(\lambda)_i$\ Component Parameters}
\begin{center}
\begin{tabular}{rccc}
$\lambda_i \; \; \; $ & $\lambda_{i,j}$ & $\gamma_{i,j}$ & $b_{i,j}$ \\  \hline
6.2 \micron  & 5.250 & 0.0300 & 0.22700 \\
             & 5.700 & 0.0400 & 0.11400 \\
             & 6.220 & 0.0284 & 0.53600 \\
             & 6.690 & 0.0700 & 0.12300 \\  \hline
7.6 \micron  & 7.417 & 0.1260 & 0.24000 \\
             & 7.598 & 0.0440 & 0.38900 \\
             & 7.850 & 0.0530 & 0.37100 \\  \hline
8.6 \micron  & 8.330 & 0.0520 & 0.18900 \\
             & 8.610 & 0.0390 & 0.81100 \\  \hline
11.3 \micron & 11.23 & 0.0100 & 0.07320 \\
             & 11.30 & 0.0290 & 0.33500 \\
             & 11.99 & 0.0500 & 0.34900 \\
             & 12.61 & 0.0435 & 0.23400 \\
             & 13.60 & 0.0200 & 0.00444 \\
             & 14.19 & 0.0250 & 0.00480 \\  \hline
\end{tabular}
\end{center} \label{tab:fractions}
\end{table}

\clearpage
\bibliography{spitzer_bump3}

\end{document}